\documentclass[
     aps,
     amsmath,
     amssymb,
     nofootinbib,
     superscriptaddress,
     doublefloatfix,
     table-of-contents,
     eqsecnum,
     rmp
 ]{revtex4-2}
\setcitestyle{numbers}
\usepackage{ragged2e} 
\usepackage{enumitem}
\usepackage{subcaption}
\DeclareCaptionJustification{justified}{%
  \justifying
  \setlength{\parfillskip}{0pt plus 1fil}%
}
\makeatother
\usepackage{siunitx}
\newcolumntype{B}{@{}S[
  table-format = 1.0,
  table-number-alignment = right,
  table-text-alignment = right
]@{}|}

\newcommand{\ket}[1]{|#1\rangle}
\newcommand{\bra}[1]{\langle#1|}

\newcommand{\klin}{k_{\text{lin}}}
\newcommand{\gatenorm}[1]{\lVert\textsc{#1}\rVert_{\text{Gates}}}
\newcommand{\negl}[1]{\mathsf{negl}(#1)}
\newcommand{\ustep}[1]{\xrightarrow{#1}}

\newcommand{\Pebble}{\ustep{\textsc{P}}}

\newcommand{\SWAP}{\ustep{\textsc{SWAP}}}

\usepackage[pdftex]{graphicx}
\usepackage{tikz}
\usetikzlibrary{quantikz2}

\usepackage{xcolor}
\usepackage{mathrsfs}
\usepackage[colorlinks, breaklinks, urlcolor={blue}, linkcolor={red}, citecolor={blue}]{hyperref}
\usepackage{amsmath}
\usepackage[english]{babel}
\usepackage{booktabs}
\usepackage{xcolor}
\usepackage[dvipsnames]{xcolor}
\usepackage{amssymb}

\usepackage{type1cm}
\usepackage{caption}
\usepackage{braket}
\usepackage{url}
\usepackage[breaklinks]{hyperref}
\usepackage{multirow}
\usepackage{csquotes}
\usepackage{dutchcal} 
\usepackage{tcolorbox}
\usepackage{comment}
\usepackage{algorithm2e}
\usepackage{lmodern, tikzpeople} 

\usetikzlibrary{positioning, shapes.multipart}
\usetikzlibrary{fit,backgrounds}
\usetikzlibrary{matrix,shapes,arrows,positioning,chains, calc}
\usetikzlibrary{arrows.meta, positioning, backgrounds, fit, shapes.geometric, calc}
\definecolor{classicblue}{RGB}{0, 70, 200}   
\definecolor{quantumred} {RGB}{210, 0, 30}  
\definecolor{arrowblue}  {RGB}{0, 160, 255}
\definecolor{boxorange}  {RGB}{210, 55, 0} 
\usepackage{listings}
\DeclareMathSymbol{\ast}{\mathbin}{symbols}{"03}

\usepackage{pifont}

\usepackage{amsthm}

\newtheorem{thm}{Theorem}

\newtheorem{cor}[thm]{Corollary}
\newtheorem{defn}[thm]{Definition}

\usepackage{listofitems}
\newcommand\cycle[2][\,]{%
  \readlist\thecycle{#2}%
  (\foreachitem\i\in\thecycle{\ifnum\icnt=1\else#1\fi\i})%
}

\def\Gopi#1{\noindent
\textcolor{Red}
{\textsc{(Gopi's comment:}
\textsf{#1})}}

\def\Mia#1{\noindent
\textcolor{ForestGreen}
{\textsc{(Mia's comment:}
\textsf{#1})}}

\def\Tom#1{\noindent
\textcolor{Magenta}
{\textsc{(Tom's comment:}
\textsf{#1})}}

\begin{document}

\title{A quantum algorithm for one-shot signatures}
\author{Gopikrishnan Muraleedharan}
\email[Author to whom any correspondence should be addressed: ]{gopi@btq.tech}
\affiliation{BTQ Technologies, 16-104 555 Burrard Street, Vancouver BC, V7X 1M8 Canada}

\author{Minh Thuy Truc Pham}
\affiliation{BTQ Technologies, 16-104 555 Burrard Street, Vancouver BC, V7X 1M8 Canada}
\email[]{minh@btq.tech}

\author{Vir Pathak}
\affiliation{BTQ Technologies, 16-104 555 Burrard Street, Vancouver BC, V7X 1M8 Canada}
\email[]{vir@btq.tech}

\author{Thomas Gardner}
\affiliation{BTQ Technologies, 16-104 555 Burrard Street, Vancouver BC, V7X 1M8 Canada}
\email[]{tom@btq.tech}

\author{Chuanqi Zhang}
\affiliation{BTQ Technologies, 16-104 555 Burrard Street, Vancouver BC, V7X 1M8 Canada}
\email[]{chuanqi@btq.tech}

\author{Samuel Crew}
\affiliation{QPerfect, 23 Rue du Loess, 67200 Strasbourg, France}

\author{Tom Hartweg}
\affiliation{QPerfect, 23 Rue du Loess, 67200 Strasbourg, France}

\author{Adrian Aasen}
\affiliation{QPerfect, 23 Rue du Loess, 67200 Strasbourg, France}

\author{Guido Masella}
\affiliation{QPerfect, 23 Rue du Loess, 67200 Strasbourg, France}

\author{Vase Moeini}
\affiliation{QPerfect, 23 Rue du Loess, 67200 Strasbourg, France}

\author{Asier Pi\~neiro Orioli}
\affiliation{QPerfect, 23 Rue du Loess, 67200 Strasbourg, France}

\author{Hugo Perrin}
\affiliation{QPerfect, 23 Rue du Loess, 67200 Strasbourg, France}

\author{Gavin K. Brennen}
\affiliation{BTQ Technologies, 16-104 555 Burrard Street, Vancouver BC, V7X 1M8 Canada}
\email[]{gbrennen@btq.tech}


\begin{abstract}
 We provide a pre-obfuscation circuit-level implementation of an efficient one shot signature scheme, which has known applications to delegated signatures, secured token transfer, and publicly verifiable randomness. The algorithm consists of two stages: a key generation stage where a classical public key/quantum secret key pair is produced, and a signing stage where the quantum secret key is processed with a message string to produce a classical signature. There is no algorithmic error in the construction and the signed message can be efficiently checked by a classical verifier. Our scheme works by preparing a superposition over elements of a random affine coset determined by the output of a puncturable pseudorandom function, together with a circuit that tests coset membership. The logical qubit number scales like $\Theta( \kappa\log(r) + n + l)$ and the gate complexity scales like $\Theta(n^3 + nl)$, where  $r$ is the public key size, $n+l$ is the signature size, $l$ is the message size, and $\kappa = \Omega(n)$ is the cryptographic security parameter. We provide explicit qubit and gate counts for varying $n$ and identify the circuit components where obfuscation would be required for security against classical and quantum polynomial time attacks. 
\end{abstract}

\maketitle


\section{Introduction}

The study of quantum algorithms is frequently framed within the context of computational complexity, where seminal protocols, such as Grover's algorithm~\cite{grover1996} and Shor’s algorithm for integer factorization and breaking Elliptic Curve Cryptography (ECC)~\cite{shor1994, shor1999}, demonstrate the capacity to efficiently solve classically intractable problems. Consequently, the impact of quantum computing on modern cryptography is mainly viewed as an adversarial threat to existing public-key infrastructure, rather than a resource for building fundamentally new security models. This perceived vulnerability has catalyzed the ongoing global transition into post-quantum cryptography~\cite{NIST_PQC}. However, this perspective overlooks the constructive potential of quantum mechanics. Beyond merely undermining classical encryption, quantum computing provides a robust foundation for novel cryptographic design, enabling functionalities and security guarantees that are simply impossible within a purely classical framework.

This constructive paradigm is best exemplified by ``local quantum cryptography'' (also referred to as hybrid quantum cryptography), an emerging research field that restricts operations entirely to local quantum computation and classical communication. Crucially, this framework eliminates the need for a quantum internet. Network communication between parties remains classical, functioning identically to traditional cryptographic protocols. The innovation instead lies at the endpoints by replacing classical hardware at specific local nodes with quantum technologies. By generating and storing cryptographic assets as quantum states locally, the system drastically enhances security capabilities without requiring any alterations to the existing classical communication infrastructure. Although the work initially attracted less attention than quantum communication based cryptography, the notion has since become increasingly influential due to its useful applications and its foundational role in local quantum cryptography.

Specifically, this regime operates under the premise that while communication channels between parties remain entirely classical, the participating entities leverage local quantum states to secure and process data. By exploiting the physical constraints of quantum information, most notably the no-cloning theorem~\cite{wootters1982single} and measurement-induced state collapse, cryptographic capabilities can be physically bound to a quantum system. Slightly shifting the focus, one of the most fundamental building blocks in cryptography is the digital signature, whereby a message is signed using a secret key verifiable by anyone holding that same secret key in the symmetric setting, or by anyone holding the corresponding public key in the asymmetric setting. Many variants of digital signatures have since been proposed, each extending the base primitive with additional properties to meet the demands of specific applications. In 2020, Amos, Georgiou, Kiayias, and Zhandry introduced the concept of One-Shot Signatures (OSS)~\cite{one-shot}, a primitive that can be seen as one such variant, further extending the classical notion of digital signatures into the quantum realm, which can be marked as the foundational work for local quantum cryptography. In the context of one-shot signatures, a user's signing key is encoded directly into a fragile quantum state. The act of executing the signature algorithm necessitates a measurement that inherently and irreversibly consumes the key. This physical enforcement, together with a classical scheme  that hides the program that produces the quantum state, guarantees unclonability and ensures strict single-use policies entirely at the local level, eliminating the threat of key duplication and providing a fundamentally new primitive for secure digital delegation.

Nevertheless, the original paper~\cite{one-shot} left two major gaps: it did not present an explicit construction of the so-called equivocal hash function, a key building block of the scheme, and it contained a critical flaw in its security proof. The year 2025 marked a turning point, with several breakthroughs addressing these problems. The works by Shmueli and Zhandry~\cite{sz25} and~\cite{SZ25b} repaired the proof and provided the first concrete construction of an OSS. Later that year, Huang and Vaikuntanathan~\cite{long-message-one-shot} strengthened that construction by extending the scheme’s functionality from signing a single bit to signing longer messages.
Additional follow-up research further expanded the landscape, revealing new applications of the one-shot framework across multiple cryptographic settings. Those aforementioned works, however, are at the level of a formal protocol without providing an explicit quantum algorithm or circuit representation and resource analysis.

In this work, we give complete quantum circuit level implementations for the algorithmic components of One-Shot Signatures. The quantum key generation involves sampling a uniform superposition over states from a random affine subspace. The key component is a pseudorandom function which we implement using a pseudorandom generator called within a Goldreich-Goldwasser-Micali (GGM) construction \cite{GGM84} together with an affine subspace sampler using a novel quantum subroutine based on the Bruhat decomposition. We also introduce a simpler algorithm for signing $l$-bit messages using a global measurement and a single translation using one multi-control phase gate instead of $l$ instances thereof \cite{long-message-one-shot}.

 OSS can be thought of as part of a broader movement known as \textit{uncloneable cryptography}, where the goal is to leverage the no-cloning theorem in order to achieve cryptographic constructions which are impossible via classical computation alone. A popular technique enabling many such constructions comes from the notion of \textit{hidden subspaces}. Such schemes depend on the ability to find points in some hidden subspace given oracles. Using this technique, primitives such as public key quantum money (with classical communication) \cite{AC12}, and tokenized signatures \cite{BenDavid2023quantumtokens} became within reach. Coladangelo, Liu, Liu, and Zhandry \cite{CLLZ22} generalize this technique to \textit{hidden cosets} and show how to obtain similar constructions that are \textit{computationally secure} in a plain model (as opposed to a classical oracle model) via obfuscated coset membership checking circuits instead of oracles. The construction of OSS we implement relies directly on such principles. With it, we can obtain a simple construction of quantum money using just classical communication, as in \cite{AC12}. Furthermore, OSS yields a straightforward construction of \textit{quantum lightning}. Introduced by Zhandry, quantum lightning further strengthens quantum money by requiring that even the party who devised the original state cannot produce two valid copies; subsequent works have exploited the structural connection between OSS and quantum monetary systems to advance this program~\cite{Shm22,sz25}. This connection also motivated a concrete line of work on \emph{quantum payment schemes}. Coladangelo and Sattath~\cite{coladangelo2020quantum} showed that quantum money can replace a public ledger to address blockchain scalability. Building on this, Sattath~\cite{sattath2022prudent} used OSS as the central building block for \emph{quantum prudent contracts}: since double-spending is physically impossible by the one-shot property, no blockchain or mining is required, and transactions are locally verifiable with no throughput limit. A more recent development inspired by the creation of OSS is \emph{quantum fire}~\cite{cakan2025public, casper2026publicly, bostanci2025general}, where quantum states are simultaneously clonable and untelegraphable. Using quantum fire, with OSS as a building block, one can obtain publicly certifiable randomness and trustless random beacons~\cite{casper2026publicly}.

The security of most existing OSS constructions relies on the assumption of sub-exponentially secure Indistinguishability Obfuscation (iO). Broadly, \textit{program obfuscation} can be considered as a ``compiler'' that scrambles a program , making it difficult to reverse engineer while preserving its original functionality. In recent years, obfuscation has become an active research topic, with the primary goal of preventing attacks by obscuring the inner workings, logic, and structure of a program. Some techniques further attempt to hide input/output dependencies. Early research established that Virtual Black Box (VBB) obfuscation, the ideal standard where an obfuscated program is as opaque as a black box, is generally impossible for all circuits~\cite{iO}. Consequently, the field shifted toward iO, which guarantees that the obfuscations of any two functionally equivalent circuits are computationally indistinguishable. Approaches to constructing iO can be divided into two primary categories: quantum circuit obfuscation and post-quantum iO. The first category utilizes dummy gates, inverse gates, and mathematically equivalent ``cloaked'' gates to hide circuit logic~\cite{obfusqate,Opaque}. This approach generally relies on structural complexity and heuristic techniques rather than formal mathematical proofs of hardness; its main objective is to resist reverse engineering. However, a recent work has demonstrated attacks against this technique using ML-based algorithms~\cite{attackOpaque}. The second category, post-quantum iO~\cite{PrO,diamondiO,pviO,quasilineriO}, is built primarily upon the hardness of Learning With Errors (LWE) problems or mathematical proofs of equivalence. While this direction offers solid security proofs based on hardness assumptions, it focuses mainly on demonstrating the existence of iO schemes and remains largely impractical. Although state-of-the-art work has achieved a quasilinear blow-up compared to the size of the original circuit~\cite{quasilineriO}, this overhead remains too large for our OSS implementation. Therefore, in this work, we focus on implementing the unobfuscated version of OSS. It is important to note that the fundamental building block of OSS is the Puncturable Pseudorandom Function (PPRF); consequently, one can naturally integrate our work with future practical iO schemes.

The remainder of this paper is organized as follows. Section~\ref{sec:background} establishes the theoretical background, formalizing the definition of one-shot signatures, reviewing the foundational SZ25 protocol, introducing our streamlined construction, and discussing practical applications. In Section~\ref{sec:algorithm}, we detail our concrete algorithmic implementation, providing explicit quantum circuit constructions and workflows for key generation, multi-bit signing, and classical verification. Section~\ref{sec:resources} presents a comprehensive logical resource analysis, detailing the qubit requirements and gate complexities of our proposed circuits. Finally, Section~\ref{sec:security} analyzes the concrete security of our implementation, including parameter selection, potential attack vectors, and a detailed discussion on the necessary classical and quantum obfuscation layers, before concluding the paper in Section~\ref{sec:conclusions}.

\section{Background}
\label{sec:background}

One can think of a One-Shot Signature (OSS) scheme as an asymmetric  digital signature scheme satisfying the \emph{one-shot property}. Concretely, an entity that has a quantum computer generates a pair consisting of a quantum secret key and a corresponding classical public key. Because the secret key is a quantum state intended for a single use, reusing it is physically impossible: once used to sign a message, the state collapses upon measurement and the measurement output is the classical signature; hence, the key is irreversibly destroyed in the process. This physical destruction ensures that the key cannot be reused, copied, or duplicated after signing. Crucially, this setting requires only a local quantum computer on the signer's side; all other parties remain entirely classical, and all  communication is performed over classical channels. This is precisely why several previous works have considered OSS as a foundational primitive for a hybrid approach that keeps both the classical and quantum components, or equivalently, for \emph{local quantum cryptography}.

This stands in sharp contrast to any purely classical approach: a classical approach to attaining the \emph{one-shot property} would require some form of global consensus, such as a distributed ledger, e.g. a blockchain, or a trusted bulletin board, to enforce it externally, which significantly slows verification and introduces additional trust and scalability overheads. By contrast, when the secret key is stored as a quantum state, the one-shot property is enforced physically, making it possible to implement OSS with constant classical communication complexity, as verification requires only a direct check of the classical signature against the classical public key. 
With this intuition in place, a formal definition of the OSS scheme is given below. In the following we make use of the notation $\negl{x}$ to denote the class of functions that falls off faster than the inverse of any polynomial \cite{katz2007}.


\begin{defn}
    A one shot signature scheme ($\mathsf{OSS}$) with security parameter $\lambda$ comprises the following algorithms.
    \begin{itemize}
        \item $crs \leftarrow \textsc{Setup}(1^{\lambda})$. A classical probabilistic polynomial-time routine that samples a classical common reference string.
         The intention is that a trusted third-party runs the routine and publicizes the result.
        This defines a consistent operating environment for all interested parties.

        \item $(pk, \ket{sk}) \leftarrow \textsc{KeyGen}(crs) $. A quantum polynomial-time algorithm that samples a key pair with classical public key, $pk$, and quantum secret key, $\ket{sk}$.
        
        \item $\sigma \leftarrow \textsc{Sign}(crs, \ket{sk}, m)$. A quantum polynomial-time algorithm that samples a signature, $\sigma$, for any message $m \in \mathcal{M}_\lambda$.
        
        \item $\textsc{Verify}(crs,pk, m, \sigma) \in \{0, 1\}$. A classical deterministic polynomial-time  algorithm that verifies the message--signature pair, $(m, \sigma)$. 
    \end{itemize}
    The algorithms satisfy the following Correctness and Security properties for a negligible function $\negl{\cdot}$:
    \begin{itemize}
        \item \textbf{Correctness}: 
        \begin{align}
            \Pr\left[
            \textsc{Verify}(crs,pk, m, \sigma)=1
            \;:\;
            \begin{array}{l}
            crs \leftarrow \textsc{Setup}(1^\lambda) \\
            (pk, \ket{sk} \leftarrow \textsc{KeyGen}(crs) \\
            \sigma_m \leftarrow \textsc{Sign}(crs, \ket{sk}, m)
            \end{array}
            \right] = 1-\negl{\lambda}.
        \end{align}
        \item \textbf{Security}: For any quantum polynomial time adversary $\mathcal{A}$ and for all $\lambda\in\mathbb{N}$
        \begin{align}
            \Pr\left[
            \begin{array}{l}
            \land (m_0, \sigma_0) \neq (m_1, \sigma_1) \\
            \land \textsc{Verify}(crs, \text{pk}, m_0, \sigma_0) = 1 \\
            \land \textsc{Verify}(crs, \text{pk}, m_1, \sigma_1) = 1
            \end{array}
            \ : \
            \begin{array}{l}
            crs \leftarrow \textsc{Setup}(1^\lambda) \\
            (\text{pk}, m_0, m_1, \sigma_0, \sigma_1) \leftarrow \mathcal{A}(crs)
            \end{array}
            \right] \leq \negl{\lambda}.
        \end{align}
    \end{itemize}
    \label{def:oss}
\end{defn}

Since this work focuses on the unobfuscated OSS construction,
we concentrate attention on the three core algorithms,
$(\textsc{KeyGen}, \textsc{Sign}, \textsc{Verify})$.


\subsection{High-level structure of the SZ25 OSS protocol}
\subsubsection{Setup}
\cite{sz25} builds the scheme above from three circuits,
$P$, $P^{-1}$, and $D$, derived from the common reference string returned by $\textsc{Setup}$.
The circuits require access to two keyed pseudorandom routines,
and so are subject to obfuscation before being broadcast.

The first is a permutable pseudorandom permutation (PPRP)
    $\Pi: \mathbb{Z}_2^n \rightarrow \mathbb{Z}_2^n$,
which for clarity we write as $\Pi(x) = (y,z)$, with $y\in\mathbb{Z}_2^r$ and $z\in\mathbb{Z}_2^{n-r}$.
The second is a puncturable pseudorandom function (PPRF),
	$F:\mathbb{Z}_2^{r} \rightarrow \text{Graff}_{\mathbb{Z}_2}(n-r,n); y \overset{F}\mapsto (A_y, b_y)$,
that produces an affine Grassmannian over $\mathbb{Z}_2$, where we denote the affine Grassmannian by the matrix--vector pair that maps $\mathbb{Z}_2^{n-r}$ into the $\mathbb{Z}_2^n$ subspace. The pseudorandom function key of size $\kappa$ is hardwired into the obfuscated circuits. If any information about the PPRF key were revealed, information about the hidden subspace associated to the public verification key is leaked which could lead to signature forgeries.

With the aforementioned setup, we can build the $P$ circuit as follows.
Say that $\Pi(x) = (y, z)$, and $F(y) = (A_y, b_y)$, then $P(x) = (y, A_yz + b_y) =: (y_x, u_x)$. We will refer to these tuple elements as the hash and the coset point. The circuit $P^{-1}: \mathbb{Z}^r_2\times\mathbb{Z}^n_2\rightarrow\mathbb{Z}^n_2\cup\{\bot\}$
is in essence the inverse of $P$ extended onto $\mathbb{Z}_2^n$.
It returns $\bot$ if the input is an invalid hash--coset pair, and the corresponding $x$ otherwise. The last circuit is essential to the signing procedure. It takes the form
$D: \mathbb{Z}^r_2\times\mathbb{Z}^n_2 \rightarrow \{0,1\}$ and tests, for a given $y$ and $u$, whether $A^T_y u = 0$.

\subsubsection{Key generation}\label{KeyGenHighLevel}
A high level view of the key generation is a program like the following.
\footnote{
	We use the notation
    $\ustep{G}$ to denote the application of a quantum gate, $G$;
    $\ustep{f}$ to denote the application of the classical function $f$, implemented as a quantum query oracle, $U_f \ket{x, y} = \ket{x, y + f(x)}$;
	and $\ustep{\textsc{Circuit}}$ to denote the application of quantum circuit $\textsc{Circuit}$.
}
\begin{align*}
	\textsc{KeyGen}:
    & \ustep{\textsc{Reset}}                          \ket{0^{n+r+k}}   \\
	& \ustep{H^{\otimes n}} |X|^{-1/2}\sum_{x\in X}   \ket{x, 0^{r+k}}  \\
	& \ustep{P}             |X|^{-1/2}\sum_{x\in X}   \ket{x, y_x, u_x} \\
	& \ustep{P^{-1}}        |X|^{-1/2}\sum_{x\in X}   \ket{0, y_x, u_x} \\
	& \ustep{\text{Measure}}           |X'|^{-1/2}\ket{0,y} \sum_{x\in X'} \ket{u_x}
\end{align*}

Here, $H$ is the Hadamard gate, and the final measurement is done on the middle $r$ qubit register, which holds the public key. The final sum is over the entire secret coset that $F$ associates with $y$.
The signing procedure does not work unless the $x$ register is disentangled.
The public key is the value that was measured, and the secret key is the
partially measured state.

\subsubsection{Signing}
A valid signature for public key, $y$, is any point in the secret coset, $C_y := b_y + \operatorname{Im}(A_y)$ where $\operatorname{Im}(A_y) = \{ A_y z : z\in\mathbb{Z}_2^{(n-r)}\}$. 
Given such a point, $u$, we interpret it as a signature for bit $b$ if $u_1 = b$.
The algorithm $\textsc{Sign}(y, \ket{sk_y}, b)$  generates this value by performing an adapted Amplitude Amplification on the secret key state, and measuring the secret key state \cite{one-shot}.

\subsubsection{Verification}
The verification algorithm is a classical, deterministic polynomial-time procedure that, given a common reference string $crs$, a public key $pk = y$, a message bit $m$, and a signature $\sigma$, ensures that the generated signature is valid. Based on the protocol's construction, the verifier utilizes the inverse function $P^{-1}$, which is accessible either as a classical oracle or instantiated via indistinguishability obfuscation within the crs. The algorithm parses the public key as $y$ and the signature as $\sigma$, outputting $1$ (accept) if and only if both of the following conditions are satisfied:
\begin{enumerate}
    \item Validity of Pre-image: The evaluation of the inverse function does not output the rejection symbol, ensuring that $P^{-1}(y, \sigma) \neq \bot$.
    \item Message Binding: The first bit of the signature $\sigma$ matches the signed message bit $m$.
\end{enumerate}

\subsection{A streamlined construction for OSS}
The construction of Shmueli and Zhandry~\cite{sz25} establishes OSS in the plain model by introducing a PPRP. At a high level, the plain model implementation uses a program $P$ built from a PPRP $\Pi$ and a PPRF $F$. The PPRP serves a structural role: it ensures that the values populating the output register during key generation were produced by a controlled, obfuscatable permutation, rather than by an unprotected uniform sampling procedure. The PPRP itself is constructed on top of the underlying PPRF via a binary tally tree, a hierarchical structure that propagates pseudorandomness through successive levels to build a full-domain permutation. The PPRFs play dual roles in this construction: There is a PPRF $F$ used directly as the pseudorandom arithmetic component, and additionally, another PPRF $F'$ used as the underlying building block from which the PPRP $\Pi$ is derived. Thus, the full construction depends on PPRF instantiation that is leveraged twice. While this construction is sound, traversing the tally tree demands substantial circuit depth and qubit overhead, making a direct implementation resource-intensive.

A key observation, however, motivates a simplification. Consider an input of length $n$ bits, so that the full input space is $\mathbb{Z}_2^n$. Since $\ket{+}^{\otimes n}$ is an eigenstate of any permutation $\Pi$, applying $\Pi$ to the uniform superposition leaves the state invariant. Concretely, whereas the original $\textsc{KeyGen}$ applies $\Pi$ to the uniform superposition before passing it to $F$ and the linear arithmetic component, our modified $\textsc{KeyGen}$ algorithm replaces this step entirely with $H^{\otimes n}$, leaving $F$ and all subsequent operations unchanged. As a consequence, only the one PPRF $F$ is now used in its direct role of producing the secret-key state; $\Pi$ is no longer required. 


The $crs$ still contains obfuscated programs for $P$, $P^{-1}$ and $D$; however, these programs are modified with respect to the original construction: the PPRP $\Pi$ inside each program is omitted while the PPRF $F$ and the linear arithmetic component remain unchanged. This modification of omitting $\Pi$ does open up an attack where a second secret key could be generated after the public key $y$ has been broadcast. An adversary who knows the public key could prepare a clone of the secret-key state by running the key generation algorithm on the initial state of the form  $H^{\otimes (n-r)}\ket{0}^{\otimes (n-r)}\ket{y}$, thereby breaking the one-shot guarantee. This attack can be foiled by wrapping the obfuscation around the Hadamard layer as well as the other programs, so that that choice of initial state would not reproduce the secret key. Including the Hadamard layer within the obfuscated circuit does not,
by itself, exclude other input state attacks. In particular, an
adversary can initialize the public-key input registers of two
evaluations in a maximally entangled state (Bell-pair) so that both evaluations produce a secret-key state
associated with the same measured public key. This attack exploits
the fact that the protocol works even if the input public key register is in the maximally mixed state. It therefore survives
obfuscation that preserves this behavior on arbitrary inputs. One possible
approach is to prepend a unitary $U$ satisfying
$U\ket{0}=\ket{0}$, thereby preserving honest key generation while
altering its behavior on other inputs. The requirements on such
a modification, and the limitations are discussed in
Section~\ref{sec:keygen-input-attacks}.


Section~\ref{sec:security} analyzes concrete attacks and
parameter choices, and identifies the additional protection
of key-generation inputs required for a secure instantiation
of the streamlined construction.

\subsection{Applications}

\begin{figure*}[t!]
    \centering
    \begin{tikzpicture}[
    >=Stealth,
    node distance=2cm,
    every node/.style={font=\small}
]

\node[font=\bfseries] (title) at (5, 0.5) {Signature Delegation};

\node[draw, thick, align=center, inner sep=4pt] (legend) at (10.5, 0.5)
    {{\color{classicblue}Classical}/{\color{quantumred}Quantum}};

\node[font=\large\bfseries\color{classicblue}] (alice-label) at (1.5, -1.2) {Alice};

\node[draw=classicblue, fill=classicblue!20, minimum width=1cm, minimum height=0.7cm,
      thick] (monitor) at (2.8, -1.0) {};
\draw[classicblue, thick] (2.3,-1.35) -- (3.3,-1.35);
\draw[classicblue, thick] (2.8,-1.35) -- (2.8,-1.5);    
\draw[classicblue, thick] (2.4,-1.5)  -- (3.2,-1.5);   
\node[draw=classicblue, fill=classicblue!20, minimum width=0.55cm, minimum height=0.8cm,
      thick, right=0.1cm of monitor] (tower) {};
\draw[classicblue] (tower.north west) ++(0.1,-0.15) rectangle ++(0.35, -0.1);
\draw[classicblue] (tower.north west) ++(0.1,-0.35) rectangle ++(0.35, -0.1);

\node[color=classicblue, align=left] (alice-gen) at (1.8, -2.2)
    {$\mathsf{PQC.Gen}(1^\lambda) \to (sk, vk)$};
\node[color=classicblue, align=left] (alice-sign) at (1.8, -2.8)
    {$\mathsf{PQC.Sign}(sk,\, y) \to \sigma$};

\node[draw=boxorange!80, fill=boxorange!15, thick,
      minimum width=4.4cm, minimum height=4.1cm,
      rounded corners=3pt] (bob-box) at (9.5, -2.4) {};

\node[font=\large\bfseries\color{quantumred}] (bob-label) at (10.0, -1.1) {Bob};

\node[draw=black!70, fill=quantumred!30, minimum width=0.9cm, minimum height=0.9cm,
      thick] (chip) at (8.3, -1.1) {};
\foreach \y in {-0.2, 0, 0.2}{
    \draw[black!60, thick] ($(chip.west)+(-0.25,\y)$) -- ($(chip.west)+(0,\y)$);
    \draw[black!60, thick] ($(chip.east)+(0,\y)$)     -- ($(chip.east)+(0.25,\y)$);
}
\foreach \x in {-0.2, 0, 0.2}{
    \draw[black!60, thick] ($(chip.north)+(\x, 0.25)$) -- ($(chip.north)+(\x,0)$);
    \draw[black!60, thick] ($(chip.south)+(\x, 0)$)   -- ($(chip.south)+(\x,-0.25)$);
}
\node[quantumred!100, font=\Large] at (chip) {$\circledast$};

\node[align=left] (bob-gen) at (9.5, -2.1)
    {$({\color{quantumred}\ket{sk}},\, y) \leftarrow {\color{quantumred}\mathsf{Gen}}\;(\mathsf{crs})$};
\node[align=left] (bob-msg) at (9.4, -2.8)
    {Choose a message $m$};
\node[align=left] (bob-sign) at (9.5, -3.4)
    {$x \leftarrow {\color{quantumred}\mathsf{Sign}}\;({\color{quantumred}\ket{sk}}, m)$};
\node[align=left] (bob-dest) at (9.2, -4.0)
    {${\color{quantumred}\ket{sk}}$ self destruct};

\draw[->, line width=2pt, color=arrowblue]
    (7.3, -2.2) -- (3.6, -2.2)
    node[midway, above, black, font=\itshape] {$y$};

\draw[->, line width=2pt, color=arrowblue]
    (3.6, -2.8) -- (7.3, -2.8)
    node[midway, below, black, font=\itshape] {$\sigma$};

\node[circle, draw=black, fill=black, minimum size=0.45cm] (head) at (9.5, -6.1) {};
\node[draw=black, fill=black, minimum width=0.85cm, minimum height=0.6cm,
      rounded corners=3pt] (body) at (9.5, -6.65) {};

\node[circle, draw=classicblue, fill=classicblue, minimum size=0.4cm]
    (badge) at (9.9, -6.7) {};
\node[white, font=\tiny\bfseries] at (9.9, -6.7) {\checkmark};

\node[font=\large\bfseries] at (10.8, -6.2) {\textbf{Verifier}};

\draw[->, line width=2pt, color=arrowblue]
    (9.5, -4.5) -- (9.5, -5.9)
    node[midway, right, black] {$(m,\, y,\, \sigma,\, x)$};

\node[color=classicblue, align=center] (ver1) at (9.5, -7.3)
    {$\mathsf{PQC.Ver}(vk,\; y,\; \sigma) \to 1/0$};
\node[color=classicblue, align=center] (ver2) at (9.5, -7.9)
    {$\mathsf{Ver}\;(\mathsf{crs},\; y,\; m,\; x) \to 1/0$};

\node[draw=black, thick, align=left, text width=7.5cm,
      inner sep=7pt] (explbox) at (2.0, -6) {
    Alice wishes to allow Bob to sign \mbox{\textbf{any} single} message of his choice on her behalf.\\[6pt]
    Alice signs $y$ using some \textbf{PQC signature scheme} and sends the resulting signature back to Bob.\\[6pt]
    \textbf{Anyone can verify} the \mbox{signature that} Bob signed on Alice's behalf.
};

\end{tikzpicture}
    \caption{Signature delegation protocol. To mitigate risks from Bob's quantum capabilities, Alice uses a \textbf{post-quantum signature scheme}, generating a \textbf{signing key ($\mathsf{PQC.sk}$) and verification key ($\mathsf{PQC.vk}$)}. The blue arrows indicate classical communication that are public.}
    \label{fig:sig-dele}
\end{figure*}
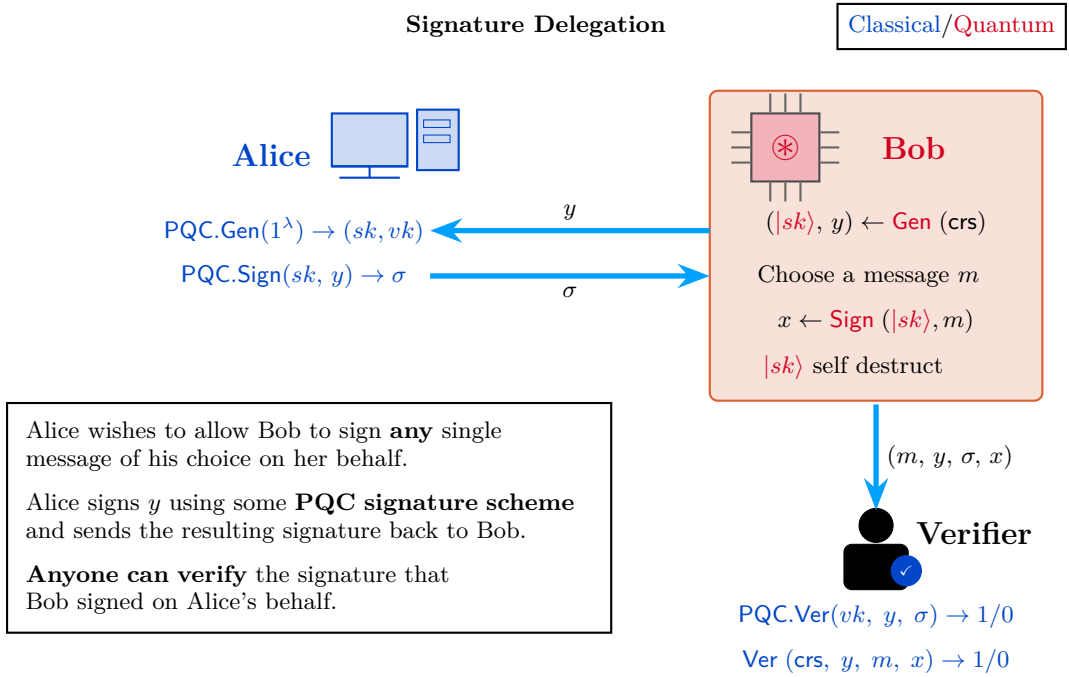

Signature delegation is one of the most important applications that motivated the creation of OSS, in which the protocol allows a party (Alice) to delegate signing authority to another party (Bob) for a single message only. This example also illustrates the continued role of Post-quantum cryptography (PQC) schemes in protecting classical parties, even within a local quantum cryptography setting. OSS is used directly in this protocol: Bob, acting as the local quantum entity, generates a quantum secret key and sends the corresponding classical public key to Alice to seek authorization. A secure PQC signature scheme is then employed, through which Alice signs Bob's public key as a means of granting that authorization. An additional party, the verifier, is also involved, responsible for checking both the validity of Alice's authorization to Bob and the authenticity of the signature produced by Bob. A more detailed description of the signature delegation protocol is given in Figure~\ref{fig:sig-dele}.

The Verifier is only considered as a passive party who verifies correctness in earlier OSS applications, which only concentrate on the interaction between Alice and Bob. One could also think about the scenario in which the Verifier has real power: the ability to control Bob's quantum abilities and the authority to accept judgements made by weaker entities like Alice. When Alice gives permission to share her data with a hospital or healthcare service, for instance, Bob, the hospital, is only permitted to access the data to do enquiries or analysis on it once. This stops Alice's privacy from being tampered with. The same principle can extend naturally to domains such as finance, banking, and government services. 

\begin{figure}[t!]
    \centering
    \begin{tikzpicture}[people/.style={minimum width=1cm}]
    \node[people, alice] (alice) {Alice};
    \node[people, bob, right=of alice] (bob) {Bob};
    \node[people, businessman, right=of bob] (char) {Charlie};
    \node[people, nun, right=of char] (dana) {Dana};

    \draw[->] ([yshift=-1cm]alice.south) coordinate (l1)--(l1-|bob) node[midway, above]{authorize};
    \draw[->] ([yshift=-1.5cm]bob.south) coordinate (l2)--(l2-|char) node[midway, above]{authorize};
    \draw[->] ([yshift=-2cm]char.south) coordinate (l3)--(l3-|dana) node[midway, above]{authorize};
  
    \end{tikzpicture}
    \caption{Signature delegation chain. The signature can easily be delegated multiple times, with Bob $\Rightarrow$ Charlie $\Rightarrow \dots \Rightarrow$ Dana and the overall signature is the entire signature chain from Alice to the final signer.}
    \label{fig:sig-dele-chain}
\end{figure}
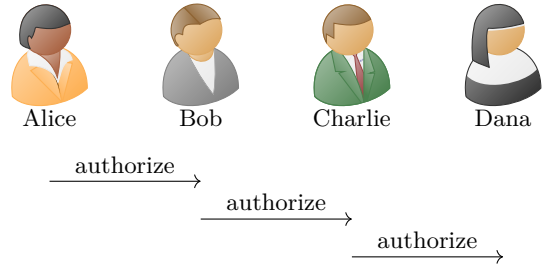

The signature delegation protocol can be naturally extended to a delegation chain. In this chain, signing authority passes sequentially through multiple parties, as shown in Figure~\ref{fig:sig-dele-chain}. Alice remains the root classical authorizer, using a PQC signature scheme to delegate authority to the first quantum party. Each subsequent quantum party plays two roles: first, by using the OSS protocol to generate a key pair and sign their message; second, by issuing a PQC authorization to the next party in the chain. The final party uses OSS solely to produce the terminal signature. The aggregate signature consists of the complete sequence of OSS signatures and PQC authorizations from all parties, which the verifier checks end-to-end. Notably, the local quantum structure is preserved: Alice and the verifier remain entirely classical. All inter-party communication occurs over classical channels. This construction has natural applications in blockchain technologies, where authorization chains are essential. The one-shot property of OSS provides a strong security guarantee in this setting. Each step is strictly single-use meaning no intermediate party can exceed their designated signing authority.

The connection between OSS and tokenized signature schemes~\cite{BenDavid2023quantumtokens} arises naturally when one observes that the OSS quantum secret key already satisfies the core requirement of a signing token: it can be  used to produce exactly one valid, publicly verifiable signature before being irreversibly consumed. In a tokenized signature scheme, a signing authority holds a classical master secret key and uses it to mint such tokens on demand. OSS strengthens this setting by removing the signing authority entirely: the quantum state itself serves as the token, generated directly by the signer with no classical master key and no trusted setup beyond a common reference string, and crucially, not even the key generator can produce two valid signatures for the same public key.

A natural application of OSS is quantum money~\cite{Wiesner83}. The quantum signing key in an OSS scheme is publicly verifiable, unclonable, and consumed upon use which are the the properties demanded of a quantum banknote. This places OSS squarely within the quantum money landscape~\cite{Shm22,sz25}, where the unforgeability of the signing key plays the role of the banknote's unclonability guarantee, and the classical verification algorithm doubles as the merchant's authenticity check, requiring no secret verification infrastructure. The connection deepens when one considers the stronger notion of quantum lightning~\cite{quantum_lightning}, where even the mint cannot produce two valid notes from the same serial number. An OSS scheme with a suitable structure, in particular, where the key generation procedure itself cannot be rewound or re-run to yield a second valid key, would satisfy an analogous guarantee. The delegation chain described above then allows a compelling monetary interpretation. Each link in the chain corresponds to a transfer of spending authority: the current keyholder, upon signing, irrevocably expends their quantum key and issues a classical post-quantum authorization designating the next recipient. The result is a chain of transfers that is locally and publicly verifiable at each step, requires no shared ledger or consensus mechanism, and is physically protected against double-spending by the one-shot property of the underlying OSS scheme. This mirrors the vision of quantum payment systems in which the unclonability of quantum states replaces the role of a blockchain, and suggests that the delegation construction may be of independent interest as a lightweight payment primitive.


OSS is a key component of quantum fire \cite{cakan2025public, casper2026publicly,bostanci2025general}.
A quantum fire state is one which is simultaneously clonable and untelegraphable. That is, given a quantum fire state, there is a procedure to ``clone" it and obtain another state that passes a public verification procedure with respect to a public serial number. However, an adversary cannot write down a classical string of polynomial length allowing another person to efficiently prepare a state passing verification. As an abstract object, quantum fire has been shown to be useful in theoretical contexts, such as a plausible method in obtaining oracle separations \cite{bostanci2025general}. Concretely, Casper, Nehoran, and Sattath \cite{casper2026publicly} constructed quantum fire in the plain model (i.e. no classical oracles) and showed how to use it to obtain publicly certifiable min-entropy. This allows for obtaining trustless random beacons, a property desired in decentralized protocols. We briefly sketch where OSS is fundamental in the quantum fire construction of \cite{casper2026publicly}. A fire state is simply a OSS signing key $\ket{sk}$. The serial number is set to be the corresponding verification key $pk$. We cannot directly clone $\ket{sk}$, since that would violate OSS security. Instead, we ask the ``cloner" to sample two public key-secret key pairs $(\ket{sk_0}, pk_0), (\ket{sk_1} pk_1)$ and subsequently obtain a signature $\sigma$ by signing $pk_0||pk_1$ using $\ket{sk}$. The cloner outputs $((pk_0, pk_1, \sigma),pk_0, \ket{sk_0})$ and $((pk_0, pk_1, \sigma), pk_1, \ket{sk_1})$. Note that we can continue to clone in the same manner: generate a pair of public key/quantum secret key pairs and subsequently sign the public keys using one of the quantum secret keys from the previous clone operation. The sequence of clones creates a path in a binary tree structure. A child node has a signature of a pair of public keys which verifies with respect to the verification key stored in the parent node. To verify, we check if the path is consistent with respect to this relationship, and if all the signatures in this path are valid. If we only allow $O(\log(\lambda))$ clones, \cite{casper2026publicly} shows this scheme is untelegraphable. The authors subsequently note that to generate any serial number useful for verification, it must possess at least some threshold amount of min-entropy. Therefore, such a serial number can serve as a min entropy source, and can be publicly verified by generating a fire state and running the verification algorithm on it. No trusted party is required, since \cite{casper2026publicly} shows that min entropy property must hold for any serial number ensuring correctness.

\section{The algorithm}
\label{sec:algorithm}

\begin{figure}[htbp]
     \centering
     \newcommand{\diagramfontsize}{\huge} 
     
     \resizebox{\textwidth}{!}{
         \begin{tikzpicture}[
     >=latex,
     font=\diagramfontsize,
     node distance=2.5em and 4.5em, 
     inputbox/.style={draw, rounded corners, inner sep=1em, align=left},
     groupbox/.style={draw, dashed, rounded corners, inner sep=1.2em}
 ]
 
 \node (keygen_circuit) {
     \begin{quantikz}[row sep=1em, column sep=1.5em] 
         \lstick{$\ket{0}^r$} & \gate{H^{\otimes r}} \gategroup[wires=5,steps=6,style={dashed,rounded corners,fill=white,inner xsep=0pt},background,label style={label position=below,anchor=north,yshift=-0.6em}]{Obfuscated} & \ctrl{4} & \qw & \qw & \qw & \ctrl{4} & \meter{} & \rstick[1]{$y=\text{pk}$} \\
         \lstick{$\ket{0}^{l}$} & \gate{H^{\otimes l}} & \qw & \qw & \gate[4]{\text{Graff}} & \qw & \qw& \qw  &  \rstick[3]{$\ket{sk_y}$} \\
         \lstick{$\ket{0}^{n-r}$} & \gate{H^{\otimes {n-r}}}& \qw & \qw & \qw & \qw & \qw & \qw &\\
         \lstick{$\ket{0}^{r}$} & \qw  & \qw & \qw & \qw & \qw  & \qw & \qw  & \\
         \lstick{$\ket{0}^{w}$} & \qw & \gate{\text{\shortstack{GGM \\ Lookup}}(\klin)} & \qw & \qw & \qw & \gate{\text{\shortstack{GGM \\ Lookup}}^\dagger(\klin)} & \qw & \rstick{$\ket{0}^{w}$}
     \end{quantikz}
 };
 \node[above=0.3em of keygen_circuit.north, font=\diagramfontsize] (keygen_label) {\textsc{KeyGen}};
 \node[inner sep=1.2em, fit=(keygen_circuit) (keygen_circuit)] (keygen_group) {};
 
 \node (keysign_circuit) [below=6em of keygen_group] 
 {
     \begin{quantikz}[row sep=1em, column sep=1.5em]
         \lstick{$\ket{\vec{y}}$} & \qw & \qw & \qw & \ctrl{4}\gategroup[wires=7,steps=8,style={dashed,rounded corners,fill=white,inner xsep=0pt},background,label style={label position=below,anchor=north,yshift=-0.6em}]{Obfuscated}  &\qw && \qw & \qw & \qw & \qw & \ctrl{4} & \qw & \qw & \rstick{$\ket{\vec{y}}$}\\
         \lstick[3]{$\ket{sk_y}$} & \meter{p} \vcw{5} \\
         & \qw & \qw & \qw && &\gate[1]{H^{\otimes n-r}} &\gate[5]{D_{p \to m}} & \gate[1]{H^{\otimes n-r}} & \qw & \qw & \qw & \qw & \meter[2]{} & \rstick[2]{$\tau$} \\
         & \qw & \qw  & \qw &  & &  \gate[1]{H^{\otimes r}} &  & \gate[1]{H^{\otimes r}} & \qw & \qw & \qw & \qw  &  & \\
         \lstick{$\ket{0}^w$} & \qw & \qw  & \qw &\gate{\text{\shortstack{GGM \\ Lookup}}(\klin)}&\qw & & \qw  & \qw & \qw & \qw & \gate{\text{\shortstack{GGM \\ Lookup}}^\dagger(\klin)} & \qw & \qw & \rstick{$\ket{0}^w$}\\
         \lstick{$\ket{0}^l$} & \gate{X} & \gate{\otimes_j X^{m_j}} & \qw & \qw &\qw & & \qw & \qw & \qw & \qw & \qw & \qw & \qw & \rstick{$\ket{\vec{p}+\vec{m}}$}\\
         \lstick{$\ket{0}^l$} & \gate{X} & \qw & \qw & \qw & \qw & \gate{H^{\otimes l}} & \qw &  \gate{H^{\otimes l}} & \qw & \qw & \qw & \qw & \qw &\meter{m}
     \end{quantikz}
 };

 \node[inner sep=1.2em, fit=(keysign_circuit)] (keysign_group) {};
 \node[above= 0.6em of keysign_group.north, font=\diagramfontsize] (keysign_label) {\textsc{Sign}};
 
 \node[right=0.6em of keysign_group.east, font=\diagramfontsize] {$\sigma = m || \tau$};
 
 
 \node[draw, rounded corners, align=center, minimum height=6.5em, text width=25em] (ggm_block) [below=42em of keysign_circuit.west, xshift=18em, anchor=west] {
     \textbf{GGM Tree Traversal $(\klin)$} \\
     (Classical Tree Search)
 };
 
 \node[draw, rounded corners, align=center, minimum height=6.5em, text width=30em] (mat_block) [right=12em of ggm_block] {
     $\tilde{\textbf{Graff}}$ \\
     $z =P_y^{-1} \cdot \left( B_y \cdot (m'+b_y^{(l)}) + \tau' + b_y^{(n)}\right)$
 };
 
 \node[draw, rounded corners, align=center, minimum height=6.5em, text width=18em] (chk1_block) [right=5.5em of mat_block, yshift=6em] { 
     \textbf{Comparator 1} \\
     Zero-Padding Check \\
     $z_{n-r \dots n} \overset{?}{=} 0^{\otimes r}$
 };
 
 \node[draw, rounded corners, align=center, minimum height=6.5em, text width=18em] (chk2_block) [right=5.5em of mat_block, yshift=-6em] { 
     \textbf{Comparator 2} \\
     Message Bit Check \\
     $m' \overset{?}{=} m$
 };
 
 \node[draw, circle, minimum size=3.5em, right=5.5em of chk1_block, yshift=-6em] (and_gate) {\textbf{AND}};
 \node[right=2.5em of and_gate, font=\diagramfontsize\bfseries] (verify_out) {Accept / Reject};

 \node[draw, dashed, rounded corners, inner sep=1.5em, fit=(ggm_block) (chk1_block) (chk2_block) (and_gate) ] (verify_group) {};
 \node[font=\diagramfontsize, fill=white, inner xsep=4pt, anchor=north, yshift=3.1em] at (verify_group.north) {$\textsc{Verify}$};
 
 \node[font=\diagramfontsize, fill=white, inner xsep=4pt, anchor=north] at (verify_group.south) {Obfuscated};
 
 \node[inputbox, left=3.5em of verify_group.west] (verify_inputs) {
     \textbf{Verification Inputs} \\[0.6em]
         $y$, $m$, $\sigma = m'||\tau'$
 };
 
 \draw[->, thick] (verify_inputs.east) -- (verify_group.west);
 \draw[->, thick] (ggm_block.east) -- node[above, font=\diagramfontsize] {Leaf Key} (mat_block.west);
 \draw[->, thick] (mat_block.east) -- node[above, font=\diagramfontsize, pos=0.3] {$z$} (chk1_block.west);
 \draw[->, thick] (chk1_block.east) -- ++(1.5em,0) |- (and_gate.135);
 \draw[->, thick] (chk2_block.east) -- ++(1.5em,0) |- (and_gate.225);
 \draw[->, thick] (and_gate.east) -- (verify_out);
 
 \path (keygen_group.west) -- (keysign_group.west) coordinate[midway] (mid_left);
 
 \node[inputbox, left=5em of mid_left, yshift=20em, anchor=east] (inputs) {
     \textbf{\diagramfontsize Inputs} \\[0.6em]
     \begin{tabular}{@{}l l@{\hspace{2.5em}}r@{}} 
     public key                   & $r$                 &   \\
     secret key length            & $n'=n+l$             &   \\
     input key for PPRF            & $\klin$    &   \\
     message length               & $l$                 &   \\
     cryptographic security parameter      & $\kappa =|\klin|$             &   \\
     workspace                    & $w$      &   \\
     message                      & $m$                 &   
     \end{tabular}
 };
 \node[font=\diagramfontsize\scshape, above=0.6em of inputs.north west, anchor=south west] {Contained in CRS};
 
 \end{tikzpicture}
  }
     \caption{Workflow diagram of the multi-bit OSS protocol. The dashed boxes represent the part of the circuits that need to be obfuscated and are distributed as classical instructions contained in CRS. Note that in {\textsc{Sign}} even though the control register $\ket{\vec{y}}$ representing the public key is classical, it should be treated as a quantum register since those gates it controls are to be obfuscated. The bottom section outlines the corresponding classical verification protocol, which efficiently evaluates signature validity using the provided CRS parameters. The binary matrix inversion inside $\tilde{\textbf{Graff}}$ operation can be easily done by reversing the classical sequence of gates used to perform binary multiplication by $A_y$.}
     \label{fig:OSS_workflow}
 \end{figure}
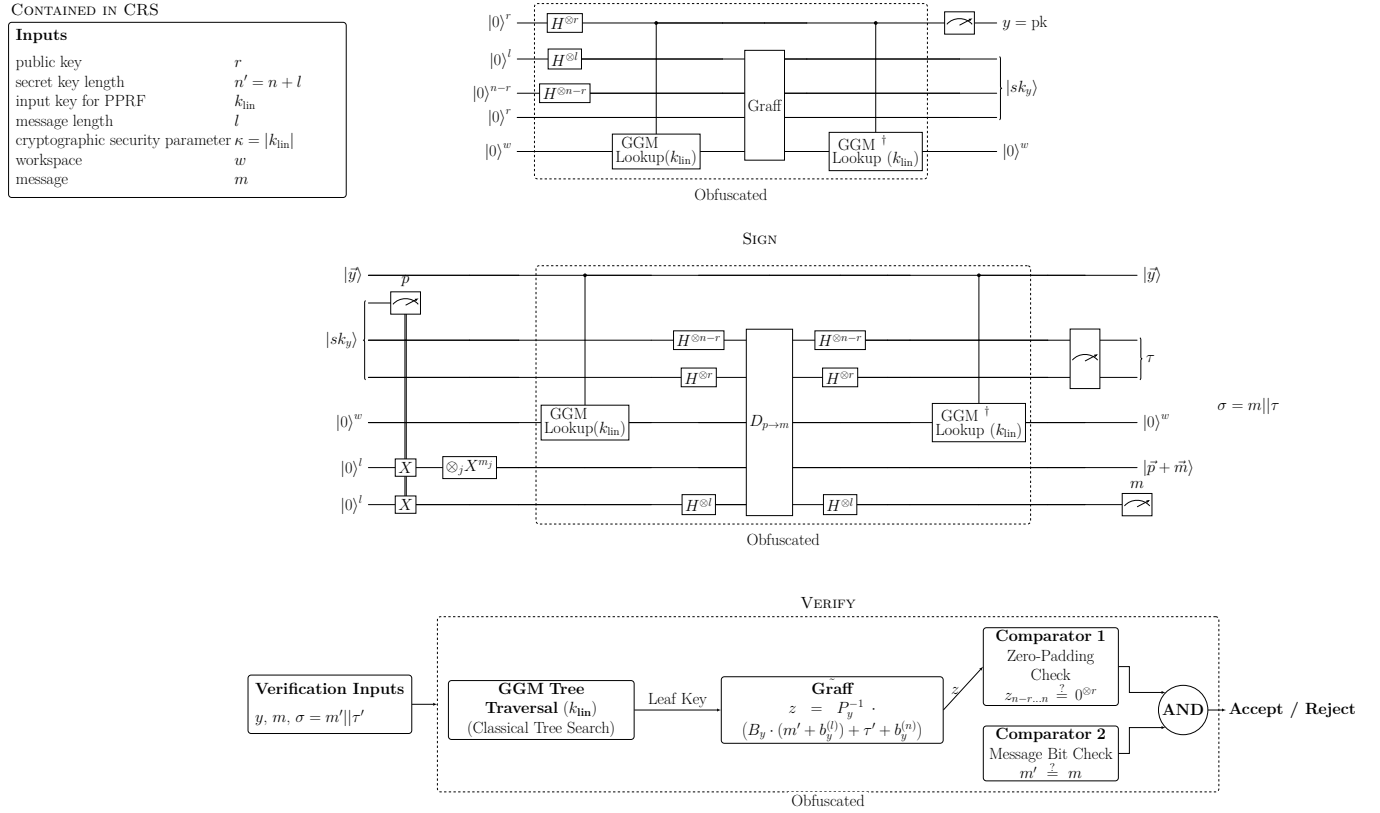

In this section, we discuss in detail our implementation of the OSS protocol providing explicit constructions of different parts of the algorithm.
The Figure~\ref{fig:OSS_workflow} shows abstract circuit diagrams for our programs.
We start with an explanation for the circuit parameters,
and follow up with detailed descriptions of each subcircuit.

\subsection{Parameters}
\label{parameters}
Let $\lambda \in \mathbb N$ be the statistical security parameter.
Recall that the OSS Construction $52$ in Shmueli and Zhandry's work~\cite{sz25} defines the following parameters in terms of $\lambda$: $n,r,k,s,d,\kappa, l$.
The parameters $n,r,k,s$ determine the sizes and dimensions of subspaces and cosets; $l = \mathcal{O}(n)$ denotes the length of the signed message; and $\kappa$, which is the size of PPRF key, determines the security level of the underlying pseudorandom function.
For most of our work, we follow Shmueli and Zhandry's recommendations in parameter instantiation, with one exception: while they recommend $s = 16\lambda$, we use $s = 2\lambda$.
This choice is the smallest value satisfying the structural requirement \cite[Lemma 32]{sz25} $s' = s - (n-r-s) > 0$,  as well as ensuring the conditions in the lemmata are satisfied. In \cite{sz25}, the constraint $s \geq 16\lambda$ is employed to rule out all possible QPT attacks but $s = 2\lambda$ is sufficient to rule out the most likely attacks described in Sec.\ref{sec:security}.

Following \cite{sz25}, we set

\begin{align}
    r = s\cdot(\lambda - 1) = 2\lambda(\lambda - 1), \quad
    n = r + \tfrac{3}{2}\cdot s = 2\lambda^2 + \lambda, \quad
    k = n,
\end{align}
and choose the PPRF key length $\kappa$ such that for $0<\delta<1$
\begin{align}
    \kappa&\geq (9 + 2\lambda(\lambda - 1) + \lambda^\delta + 2\lg(2\lambda^2+\lambda))^{1/\delta}.
\end{align}
Here, $n$ denotes the input length, $r$ the the size of the OSS public key, and we pick length of the PPRF key so that $\kappa = |\klin|$.
The parameter $s$ is the bloating factor of the dual subspace, expanding it by exactly $s$ dimensions \cite{sz25}. 
More discussion on the possible attacks we consider, security parameters, and how they are chosen can be found in Sec.~\ref{security}

\subsection{Cryptographic primitives}
A pseudorandom function (PRF) is a deterministic algorithm that takes as input
a key $k$ from a key space $\mathcal{K}$, an input $x$ from an input space $\mathcal{X}$,
and outputs $F(k, x)$, which is computationally indistinguishable from the output of a random function.
A Puncturable PRF (PPRF) is a PRF that has an extra feature that allows one to evaluate at all points except those in $S$ by ``puncturing'' a set of points $x\in S$ using a punctured key $k^S$. The holes are controlled by additional key derivation and evaluation algorithms;
technical details are in the next definition.

\begin{defn}[Puncturable PRFs~\cite{puncturedprograms13}]
A \textit{puncturable pseudorandom function} (PPRF) is a tuple of efficient algorithms $(\mathsf{F}, \mathsf{Punc}, \mathsf{Eval})$ with associated output-length function $\mathrm{poly}(\kappa)$ such that:

\begin{itemize}
    \item $\mathsf{F}: \{0,1\}^{\kappa} \times \{0,1\}^* \to \{0,1\}^{\mathrm{poly}(\kappa)}$ is a deterministic polynomial time algorithm.
    
    \item $\mathsf{Punc}(k, S)$ is a probabilistic polynomial algorithm that outputs a punctured key $k^S$ for a set of points $S \subseteq \{0,1\}^*$. 
        
    \item $\mathsf{Eval}(k^S, x)$ is a deterministic polynomial time algorithm.
\end{itemize}

\noindent\textbf{Correctness}: For any $\kappa \in \mathbb{N}$, $S \subseteq \{0,1\}^*$, $k \in \{0,1\}^{\kappa}$, $x \notin S$, and $k^S$ in the support of $\mathsf{Punc}(k, S)$, we have that $$\mathsf{Eval}(k^S, x) = \mathsf{F}(k, x).$$

\noindent\textbf{Security}: 
A PPRF function $F$ is $(f(\kappa),\negl\kappa)$-secure if for some function $f:\mathbb{N}\to \mathbb{N}$ and $\mathsf{negl}:\mathbb{N}\to [0,1]$ if for any $f(\kappa)$-time (quantum) adversary $\mathcal{A}$, the probability that $\mathcal{A}$ wins the following game is at most $\frac{1}{2}+\negl\kappa$:

\begin{itemize}
  \item $\mathcal{A}(1^\kappa)$ generates a set of points $S \subseteq \{0,1\}^*$.
  \item The challenger chooses a random key $k \leftarrow \{0,1\}^\kappa$ and computes the punctured key $k^S \leftarrow \mathsf{Punc}(k,S)$. For each $x \in S$, it also sets $y_x^0 := F(k,x)$ and samples a uniformly random string $y_x^1 \leftarrow \{0,1\}^{\mathrm{poly}(\kappa)}$. Then, the challenger chooses a random bit $b$ and gives $k\{S\}$ and $\{(x,y_x^b)\}_{x\in S}$ to $\mathcal{A}$.
  \item $\mathcal{A}$ outputs a guess $b'$ for $b$ and wins the game if and only if $b' = b$.
\end{itemize}

Even given access to the punctured key $k^S$—which allows evaluation of the PPRF at all points except $x^*\in S$—no efficient adversary can distinguish 
the outputs of the PPRF from truly random values.

\end{defn}

A classic construction for a PPRF is the GGM tree.
This construction iterates a pseudorandom generator to build a tree of keys.
Recall that a pseudorandom generator uses a short, random seed to produce a longer bit sequence that is computationally indistinguishable from a uniform random distribution.

\begin{defn}[Pseudorandom generator (PRG)]
    A pseudorandom generator (PRG) is a function $G:\{0,1\}^{\kappa}\to \{0,1\}^{\mathrm{poly}(\kappa)}$
    such that: 
    \begin{itemize}
        \item $G$ is an efficient deterministic algorithm,
        \item For any polynomial time algorithm  $\mathcal{A}$, for a random seed $s$ sampled from $\{0,1\}^{\kappa}$ and a value $r$ uniformly sampled from $\{0,1\}^{\mathrm{poly}(\kappa)}$, the distinguishing advantage of $\mathcal{A}$ between $G(s)$ and $r$ is negligible for a negligible function $\negl{\kappa}$ , i.e. 
        \begin{align}
            \big|\Pr_{s\gets\{0,1\}^{\kappa}}[\mathcal{A}(G(s))= 1] - \Pr_{r\gets\{0,1\}^{\mathrm{poly}(\kappa)}}[\mathcal{A}(r)= 1]\big|< \negl{\kappa}.
        \end{align}
    \end{itemize}
    One can also say that the uniform distribution on $\{0,1\}^{\mathrm{poly}(\kappa)}$ is 
    $\negl{\kappa}$- indistinguishable from the distribution $\{G(s)|s\gets\{0,1\}^{\kappa}\}$.
\end{defn}

Given any block cipher that is a secure pseudorandom permutation (PRP),
    a secure PRG can be constructed using standard PRP-based domain-extension methods.
For instance, one may encrypt an incrementing counter and concatenate the resulting blocks (CTR mode),
    or repeatedly encrypt the previous output block as the next input (CFB mode).

The GGM construction~\cite{GGM84} builds a PPRF,
    $\mathcal{F}: \{0,1\}^{\kappa}\times\{0,1\}^{r} \to\{0,1\}^{\kappa}$,
from a length doubling PRG, $G: \{0,1\}^\kappa \rightarrow \{0,1\}^{2\kappa}$.
The output splits into two $\kappa$-bit strings,
\begin{align}
    G(s) = \underbrace{G_0(s)}_{\kappa\text{ bits}} \, \| \, \underbrace{G_1(s)}_{\kappa\text{ bits}}
\end{align}
Thae PRF is defined as,
\begin{align}
    (k,y) \stackrel{\mathcal{F}}{\mapsto} (G_{y_{-1}}(\cdots G_{y_1}(G_{y_0}(k))\cdots))
\end{align}
The input $y \in \{0,1\}^r$ determines a path of length $r$ through the  binary tree,
with the leaves representing the PRF outputs. 

\RestyleAlgo{ruled}
\SetKwComment{Comment}{/* }{ */}
\begin{algorithm}[h!]
\caption{GGM PPRF $\mathcal{F}(\cdot,\cdot)$ from a length-doubling PRG $G(\cdot)$}\label{alg:ggm_prf}
\KwIn{$(k,y)\in\{0,1\}^{\kappa}\times\{0,1\}^{r}$}
$t\gets k$\;
\For{$i\gets 0$ \KwTo $r-1$}{
    $c=y_i$\;
    $t=G_{c}(t)$
}
\KwResult{$t$}
\end{algorithm}

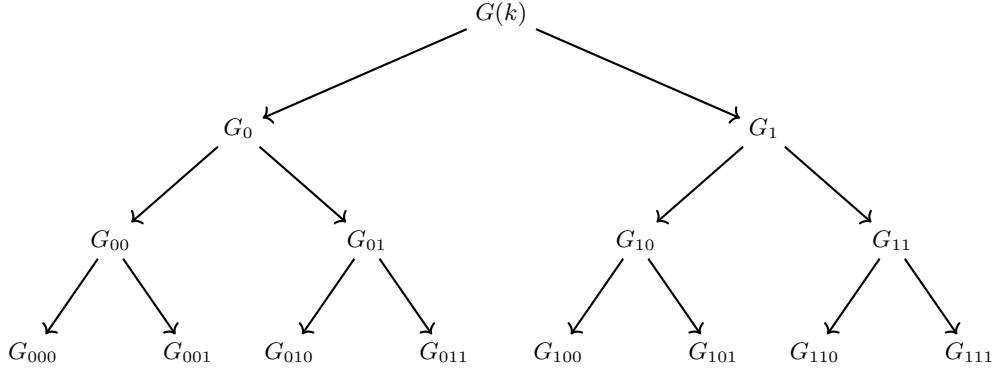
\begin{figure}[t!]
    \centering
    \begin{tikzpicture}[
    node distance=1cm and 0.2cm,
    every node/.style={align=center},
    arrow/.style={->, thick}
]

\node (root) {$G(k)$};

\node[below left=of root, xshift=-2.5cm]  (g0) {$G_0$};
\node[below right=of root, xshift=2.5cm] (g1) {$G_1$};

\node[below left=of g0, xshift=-0.8cm]  (g00) {$G_{00}$};
\node[below right=of g0, xshift=0.8cm] (g01) {$G_{01}$};

\node[below left=of g1, xshift=-0.8cm]  (g10) {$G_{10}$};
\node[below right=of g1, xshift=0.8cm] (g11) {$G_{11}$};

\node[below left=of g00]  (g000) {$G_{000}$};
\node[below right=of g00] (g001) {$G_{001}$};

\node[below left=of g01]  (g010) {$G_{010}$};
\node[below right=of g01] (g011) {$G_{011}$};

\node[below left=of g10]  (g100) {$G_{100}$};
\node[below right=of g10] (g101) {$G_{101}$};

\node[below left=of g11]  (g110) {$G_{110}$};
\node[below right=of g11] (g111) {$G_{111}$};
\draw[arrow] (root) -- (g0);
\draw[arrow] (root) -- (g1);

\draw[arrow] (g0) -- (g00);
\draw[arrow] (g0) -- (g01);

\draw[arrow] (g1) -- (g10);
\draw[arrow] (g1) -- (g11);

\draw[arrow] (g00) -- (g000);
\draw[arrow] (g00) -- (g001);

\draw[arrow] (g01) -- (g010);
\draw[arrow] (g01) -- (g011);

\draw[arrow] (g10) -- (g100);
\draw[arrow] (g10) -- (g101);

\draw[arrow] (g11) -- (g110);
\draw[arrow] (g11) -- (g111);
\end{tikzpicture}
    \caption{Sample GGM tree with $r = 3$. The output of the PRG is $G(\mathrm{k}) = G_{0} \| G_{1}$. Evaluating the GGM PPRF on input $011$ results in $\mathcal{F}(k,011) = G_{011} = G_1(G_{01})=G_1(G_1(G_0(k)))$}
    \label{fig:ggm_prf}
\end{figure}

\begin{figure}[h!]
    \centering
    \tikzset{
  box/.style={
    draw=blue,
    thick,
    rounded corners,
    inner sep=4pt,
    fit=#1
  }
}
\begin{tikzpicture}[
    node distance=1cm and 0.2cm,
    every node/.style={align=center},
    arrow/.style={->, thick},
    altarrow/.style={*-*,dashed, red}
]

\node (root) {$G(k)$};

\node[below left=of root, xshift=-2.5cm]  (g0) {$G_0$};
\node[below right=of root, xshift=2.5cm] (g1) {$G_1$};

\node[below left=of g0, xshift=-0.8cm]  (g00) {$G_{00}$};
\node[below right=of g0, xshift=0.8cm] (g01) {$G_{01}$};

\node[below left=of g1, xshift=-0.8cm]  (g10) {$G_{10}$};
\node[below right=of g1, xshift=0.8cm] (g11) {$G_{11}$};

\node[below left=of g00]  (g000) {$G_{000}$};
\node[below right=of g00] (g001) {$G_{001}$};

\node[below left=of g01]  (g010) {$G_{010}$};
\node[below right=of g01] (g011) {$G_{011}$};

\node[below left=of g10]  (g100) {$G_{100}$};
\node[below right=of g10] (g101) {$G_{101}$};

\node[below left=of g11]  (g110) {$G_{110}$};
\node[below right=of g11] (g111) {$G_{111}$};
\draw[altarrow] (root) -- (g0);
\draw[arrow] (root) -- (g1);

\draw[arrow] (g0) -- (g00);
\draw[altarrow] (g0) -- (g01);

\draw[arrow] (g1) -- (g10);
\draw[arrow] (g1) -- (g11);

\draw[arrow] (g00) -- (g000);
\draw[arrow] (g00) -- (g001);
\draw[arrow] (g01) -- (g010);
\draw[altarrow] (g01) -- (g011);
\draw[arrow] (g10) -- (g100);
\draw[arrow] (g10) -- (g101);
\draw[arrow] (g11) -- (g110);
\draw[arrow] (g11) -- (g111);

    \node[box=(g1)] {};
    \node[box=(g00)] {};
    \node[box=(g010)] {};
\end{tikzpicture}
    \caption{
        Puncturing the GGM tree at $011$.
        The puncturing key is $k^{011}= \{G_{1},G_{00},G_{010}\}$.
        The punctured path is marked with a red dashed line, and the puncturing key's nodes are in blue boxes.
    }
    \label{fig:punc}
\end{figure}
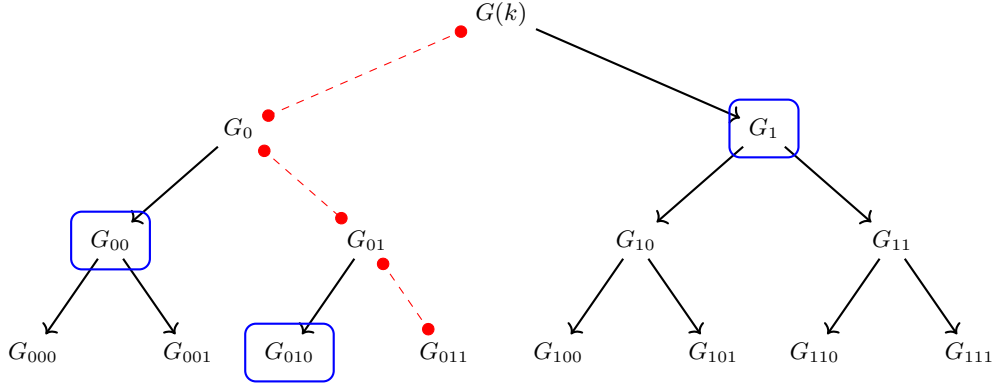

Since the GGM construction is puncturable~\cite{obfuscation}, the punctured key for a PPRF allows evaluation of the PRF tree at all points except a specific input $y^*$. 
Let $P_{y^*}$ be the path from $\mathrm{root}$ to $y^*$ and $N_{y^*}$ be the set of all neighbouring nodes in $P_{y^*}$.
The punctured key is the set of values of all the nodes in $N_{y^*}$.
These values define the roots of every subtree that doesn't contain a point in the punctured path,
and so all unpunctured points can still be evaluated with a standard GGM lookup.
Figure~\ref{fig:punc} gives a simple example of the pucnturable property of the GGM construction.
Notice that in our algorithm, it was not necessary for us to generate a punctured key for any subroutines.
However, this property is important as it is compatible with the requirement of iO. See Sec.~\ref{subsubsec:hide_k_lin} for more details.

\subsection{Setup}
Our setup is done at a higher level than \cite{sz25}.
A trusted third party samples a $\klin\in \{0,1\}^{\kappa}$ and uses it to construct a $crs$ that describes quantum programs: \textsc{KeyGen} and \textsc{Sign}, and a classical program \textsc{Verify}.
These programs are obfuscated before being publicized.
The necessary requirements for obfuscation are detailed in Sec.~\ref{sec:security}.

\subsection{KeyGen}

As seen in the Workflow diagram~\ref{fig:OSS_workflow}, the key generation circuit factors as
$\textsc{KeyGen} = H^{\otimes n+l}; F; \textsc{Measure}$.
The factor $F$ is a puncturable PRF that relates public keys to a hidden coset space,
encoded as an affine transformation ($z\mapsto A_yz + b_y$).

The GGM construction (detailed above) is a flexible recipe for building PPRFs.
However a GGM path traversal alone can not construct the desired $F$.
This is for two reasons:
1. it does not provide enough length, and 2. it does not guarantee the requisite structure. 


Our construction solves both problems.
It begins with a general quantum circuit for GGM tree path traversal.
The GGM result is used to seed a stream cipher, which solves the length problem.
The stream is passed to a novel circuit that decodes bit strings into instructions for a suitable affine transformation; this solves the structure problem.
We call these factors $\textsc{GGM.Lookup}$ and $\textsc{Graff}$, and describe their implementation in the sections that follow.

\subsubsection{Implementing the \textsc{GGM.Lookup}}
The GGM PPRF construction~\cite{GGM84} creates a binary tree of keys
by iterating a length-doubling PRG, $ G:\{0,1\}^{\kappa} \to \{0,1\}^{2\kappa}$,
along any path $p\in\{0,1\}^*$, in the pattern
\begin{align*}
      G(k) &= (k_0, k_1)           \\
    G(k_p) &= (k_{p||0}, k_{p||1})
\end{align*}
We label $k$ the root key of the tree.
The leaves define the value of the PPRF on a point $x \in \{0,1\}^r$, which defines a path through the tree from the root.

To implement this as a quantum circuit, we reduce the GGM lookup to a reversible pebble game on a tree.
Recall that reversible pebble games are an abstraction for reversible computations.\cite{bennett89}
Games are played on a directed graph with no cycles.
Every node (in the graph) maintains a counter that tracks the number of pebbles sitting on top.
The game starts with all counters at zero;
each turn adds or removes a pebble,
subject to the constraint that
	pebbles may only be added to or removed from nodes that have a pebble on all of their parents;
the game is over once a pebble is placed on some special target node.
Each pebble added to the graph corresponds to one register of space used in the computation.

The reduction is seen by specifying a pebble operator.
It's always possible to pebble the root, since the root has no parents.
This is implemented with a circuit, $\textsc{kLin}$, that writes $\klin$ into a register using CNOT gates.
The children are pebbeled by the $\textsc{NextKey}$ query oracle,
$\ket{p, k_p , 0^\kappa} \ustep{\textsc{NextKey}} \ket{p, k_p, k_q}$.
This circuit uses the value at node $p$ to select which half of the length doubling PRG to build.

The \textsc{NextKey} circuit can run in one of two modes, determined by the blockcipher's key and block size. In the case where the blocksize divides keysize in two, we initialize two blocks to $\ket{0^{\kappa/2}, 0^{\kappa/2-1}1}$, and use the point on the path to control the second qubit.
This gives $\ket{0^{\kappa/2-2}y_i0, 0^{\kappa/2-2}y_i1}$,
which is then enciphered.
Otherwise, (eg, when using the Simon cipher with $\kappa=192$) we must first allocate a separate workspace to build the key.
Sufficient key material is then copied out of the workspace into a key register, and the workspace is uncomputed.

The simplest reversible path traversal circuit caches every intermediate key.
It uses $\Theta(r)$ \textsc{NextKey} calls and $\Theta(r\kappa)$ logical qubits for a depth $r$ tree.
This is infeasible for current machines, which are bottlenecked by qubit cost. A realization is however possible by trading time complexity for space efficiency.

To proceed, we present the key registers as a comma-separated row of numerals.
Let $0$ denote a reset register, say $\ket{0^\kappa}$, $\ket{1}$ denote the GGM root ($\klin$), and successors denote the key value along the GGM path.
The gadgets above in this shorthand become
\begin{align*}
    0    &\xleftrightarrow{O} 1  \\
    n, 0 &\xleftrightarrow{P} n, n+1 \\
\end{align*}
We use subscripts to precise relevant registers (eg, $n,m,0 \ustep{P_{0,2}} n,m,n+1$) and omit them when obvious.
The pebble operation with a fixed register also creates new single-register reversible operations,
$0 \xleftrightarrow{\textsc{P}} v_i + 1$ where $v_i$ denotes the value in register $i$.
We reorder registers with a SWAP operation to simplify bookkeeping.
The SWAPs can always be optimized out and do not contribute to gate counts. 

To minimize qubit counts, we ask the question:
what is the greatest $x$ that processes of the form $0^R \xleftrightarrow{} x, 0^{R-1}$ can reach?
An adapted binary search works well.
Use $W^R_O$ to denote the optimal process on $R$ registers with a circuit $O$ that pebbles the root.
Run $W^{R-1}_O$ on the rightmost $R-1$ registers to get $0^R \ustep{W} 0, x^*_{R-1}, 0^{R-2}$.
Swap $x_{R-1}$ into register zero, and run $W^{R-1}_{P_0}$ on the rightmost $R-1$ registers, with the zero pebble set to use $P_0$.
This gives $x_{2(R-1)} , x_{R-1} , 0^{R-2}$ after swapping.
The $x_{R-1}, 0^{R-2}$ can be uncomputed by running $W$ in reverse.
That is $W^R_O = W^{R-1}_O ; \text{SWAP} ; W^{R-1}_{P_{R-1}} ; \text{SWAP}; W^{R-1}_O{}^\dagger$ for $R>1$ and $W^1_O=O$.
The $R=3$ case shows the structure:
\begin{align*}
    \ustep{\textsc{Reset}} 0, 0, 0
    \ustep{O}              0, 0, 1
    \SWAP                  0, 1, 0
    \Pebble                0, 1, 2
    \SWAP                  0, 2, 1
    \ustep{O}              0, 2, 0
    \SWAP                  2, 0, 0
    \ustep{W}              2, 4, 0
    \SWAP                  4, 2, 0
    \ustep{W^\dagger}      4, 0, 0
\end{align*}

This gives a lower bound on how far we can place a single pebble:
$x^*_R = 2x^*_{R-1}$,
and solving the recurrence yields
$x^*_R = 2^R$.
Iterating the process on the remaining free registers reaches a maximum depth of $2^R-1$.

The GGM Lookup routine in our OSS algorithm requires traversing a path to depth $r+1$,
where $r$ is the length of the public key.
This means at least $R = \lceil \lg(r + 2) \rceil$ registers are required to reach that depth.
Moreover, since each $W$ splits into three subproblems with half the problem size,
the algorithm must make $\Theta(r^{\lg(3)})$ calls to \textsc{NextKey}.

We note that this approach is in essence Bennett's simulation technique from~\cite{bennett89} with $m=2$.

\subsubsection{Implementing the PRG}
The GGM circuit builder works for any reversible implementation of a block cipher,
$\ket{k, x} \ustep{\textsc{Cipher}} \ket{k, \textsc{Cipher}_k (x)}$.
As noted above, the length-doubling PRG can be implemented by a block cipher run in a suitable streaming mode.
Because logical qubits are the dominant resource bottleneck at present,
we have focused our attention on lightweight block ciphers.
Two that seem promising are Present~\cite{present} and Simon~\cite{SIMON}.
Since Simon presents far more block and key size configurations,
we have expressed all concrete resource counts below assuming a Simon implementation
with a suitable choice for $\kappa$.

PRESENT~\cite{present} is a light weight block cipher that encrypts 64-bit data blocks with a key length of 80 or 128 bits. PRESENT follows the Substitution Permutation Network (SPN) method and has 31 rounds;
each round consists of three steps: AddRoundKey, Sbox, and Permutation.
An efficient implementation of PRESENT on quantum computers is introduced in ~\cite{present_implementation}.

SIMON~\cite{SIMON} is a family of the lightweight block ciphers introduced by
the U.S National Security Agency in 2013 that offers a wider range of block
sizes, key sizes, and rounds.
\cite{grover_on_SIMON} developed a reversible quantum circuit for all variants of SIMON.
Their implementation produces the correct ciphertext, but does not unwind the key schedule.
Our implementation unwinds the key schedule and leaves the input key intact;
this is essential for an efficient GGM path traversal.
Gate counts for our implementation are tabulated in Table~\ref{tab:simon}.

\begin{table}[h!]
\caption{Gate counts for the \textsc{Simon} block cipher $\ket{k, x} \ustep{\textsc{Simon}} \ket{k, \text{Simon}_k(x)}$ with key length $\kappa$ and blocksize \text{bk}.}
\label{tab:simon}
\begin{tabular}{ r| S | S | S | S | S | S}
    $\kappa$ &    64 &   96 &   128 &   144 &   192 &   256 \\
    \text{bk} &   32 &   48 &    64 &    96 &   128 &   128 \\
    \hline\hline
    X        &   812 & 1442 &  2438 &  4742 &  8252 &  8500 \\
    CX       &  4608 & 7872 & 13056 & 14976 & 25728 & 44032 \\
    CCX      &   512 &  864 &  1408 &  2592 &  4416 &  4608 \\
\end{tabular}
\end{table}

\subsubsection{Uniform coset generation using Bruhat decomposition}
Proceeding with the algorithmic workflow, the subsequent step involves generating the coset.
As in the \cite{sz25} construction, our implementation uses the PPRF $F$ to sample a random coset $C_y$ defined by a full-column rank matrix $A_y\in\mathbb{Z}_2^{n\times (n-r)}$ and a random vector $b_y$ $\in$ $\mathbb{Z}_2^n$.
We construct this PPRF $F$ by leveraging a GGM tree traversal to derive a leaf key, followed by repeated applications of a block cipher. Specifically, the leaf key is computed as $k_y = \mathcal{F}(\klin, y)$, which serves as the secret key for a block cipher operating in ciphertext feedback mode to expand the output to the desired length. The pseudorandom bit string output by $F(\klin, y)$ is subsequently used in conjunction with Bruhat decomposition to generate a matrix-vector pair $(A_y, b_y)$, which defines a random coset. 

\begin{align}
F: \{0,1\}^{\kappa} \times \{0,1\}^{r} &\longrightarrow \{0,1\}^* \nonumber \\
(\klin, y) &\xmapsto{\quad \mathcal{F} \quad} k_y = \mathcal{F}(\klin,y) \nonumber \\
&\xmapsto{\quad E \quad} E_{k_y}(\text{iv}) \parallel E_{k_y}(E_{k_y}(\text{iv})) \parallel \dots
\end{align}
where $\text{iv}$ is the initialization vector of the block cipher.  
The columns of $A_y$ span an $(n-r)$-dimensional linear subspace $\operatorname{Im}(A_y) = \{ A_y z : z\in\mathbb{Z}_2^{n-r}\}$. A coset $C_y$ of $\operatorname{Im}(A_y)$ is obtained by adding a vector $b_y$ to every element of $\operatorname{Im}(A_y)$, thereby shifting the entire subspace by $b_y$.  In our setting, $C_y$ serves as the affine subspace associated with $y$:
\begin{align}
   C_y := b_y + \operatorname{Im}(A_y),
    \qquad
    \operatorname{Im}(A_y) = \{ A_y z : z\in\mathbb{Z}_2^{n-r}\}.
\end{align}

One subtle part left is that the output of the  GGM PPRF $F$ construction above is a pseudorandom bit string. Here, we describe how to form a random coset from $F(\klin,y)$ using the Bruhat decomposition.    
\begin{enumerate}[label=\alph*.]
    \item Sampling a random coset
    
    One can think of the coset $C_y$ as an element of the affine Grassmannians $\text{Graff}_{\mathbb{Z}_2}(n-r,n)$, the set of all $(n-r)$-dimensional affine subspaces of $\mathbb{Z}_2^n$, represented concretely by the pair $(A_y,b_y)$ where: 
    \begin{align}
      \text{Graff}_{\mathbb{Z}_2}(n-r, n) := \{(\mathcal{L}, b) : \mathcal{L} \in \mathrm{Gr}(n-r,n)(\mathbb{Z}_2),\ b \in \mathbb{Z}_2^n\}.
    \end{align}
        The set of all $n-r$-dimensional subspaces of an $n$-dimensional vector space over a field $\mathbb{Z}_2$, denoted by $Gr(n-r,n)(\mathbb{Z}_2)$, is called the Grassmannian. 
        The affine Grassmannian  $\text{Graff}_{\mathbb{Z}_2}(n-r,n)$ extends this set by allowing subspaces to be shifted by $b_y \in \mathbb{Z}_2^n$. Hence, to sample a random coset $C_y \in \text{Graff}_{\mathbb{Z}_2}(n-r,n)$, we first sample a random element $\operatorname{Im}(A_y)$ in $\mathrm{Gr}(n-r,n)(\mathbb{Z}_2)$, and sample a random vector $b_y$.

    Note that any binary matrix $A_y \in \mathbb{Z}_2^{\,n \times (n-r)}$ can always be written in the form
    \begin{align}
    A_y = P_y \begin{bmatrix} I_{n-r} \\ R_y \end{bmatrix} S_y \label{eq:general_form_ay}
    \end{align}
    for some matrices $P_y \in \mathrm{GL}_n(\mathbb{Z}_2)$, $S_y \in \mathrm{GL}_{n-r}(\mathbb{Z}_2)$, and $R_y \in \mathbb{Z}_2^{\, r \times n-r}$.
    Its columns span a $(n-r)$-dimensional subspace of the $n$-dimensional vector space over the field~$\mathbb{Z}_2$ (with $r < n$).

    Consider any invertible matrix $P_y \in \mathrm{GL}_n(\mathbb{Z}_2)$ and a fixed $(n-r)$-dimensional subspace $\operatorname{Im}(A_y) \subseteq \mathbb{Z}_2^n$ represented by the columns of a full-column-rank matrix $A_y$. Multiplying this matrix by $P_y$ simply produces another matrix whose columns span a new $(n-r)$-dimensional subspace $P\operatorname{Im}(A_y)$. Moreover, every $(n-r)$-dimensional subspace of $\mathbb{Z}_2^n$ can be expressed in this way: for any such subspace $\mathcal{L}'$, there exists an invertible matrix $P_y$ for which
    \begin{align}
        P_y\operatorname{Im}(A_y) = \mathcal{L}'.
    \end{align}
    A convenient choice of reference subspace is the canonical one
    \begin{align}
        \mathbb{Z}_2^{\,n-r} := \mathrm{span}\{e_1, \dots, e_{n-r}\},
    \end{align}
    and any other $(n-r)$-dimensional subspace of $\mathbb{Z}_2^n$ can be written as $P_y \mathbb{Z}_2^{\,n-r}$ for some $P_y \in \mathrm{GL}_n(\mathbb{Z}_2)$. Hence, multiplying the canonical matrix by different invertible matrices $\mathrm{GL}_n(\mathbb{Z}_2)$ generates all such subspaces.
    This also shows that the specific matrices $R$ and $S$ appearing in the decomposition~\ref{eq:general_form_ay} are not important for sampling a random subspace. Thus, we may set $R = 0$ and $S = I_{n-r}$, yielding the simplified form
    \begin{align}
        A_y = P_y\begin{bmatrix}
                        I_{n-r}\\[2pt]
                    0 \end{bmatrix}.
    \end{align}

The problem of sampling a random linear subspace is now reduced to sampling a matrix $P_y\in \mathrm{GL}_n(\mathbb{Z}_2)$. We employ the Bruhat decomposition to uniformly sample invertible binary matrices; unlike other methods, it directly maps the generated random bits to the exact sequence of gates to be applied. This also makes it easier to implement $A_y^{-1}$ or $A_y^{T}$ .

\item Bruhat Decomposition

For $\mathrm{GL}_n(\mathbb{Z}_2)$, let $B$ be the subgroup of $\mathrm{GL}_n(\mathbb{Z}_2)$ consisting of all upper triangular invertible matrices and $S_n$ be the group of all permutation matrices. The Bruhat decomposition~\cite{bruhat_decomposion} then takes the form
\begin{align}
    \mathrm{GL}_n(\mathbb{Z}_2) = \bigsqcup_{\pi \in S_n} B \, \pi \, B.
\end{align}
Thus every matrix $P_y \in \mathrm{GL}_n(\mathbb{Z}_2)$ can be written as
\begin{align}
    P_y = U_1 \, S_\pi \, U_2,
\end{align}
weith $U_1, U_2 \in B$ and $S_\pi$ a permutation matrix.
\end{enumerate}

\RestyleAlgo{ruled}
\SetKwComment{Comment}{/* }{ */}
\begin{algorithm}[h!]
\caption{Sample uniform element of $\text{Gr}(n-r,n)(\mathbb{Z}_2)$ using Bruhat Decomposition}\label{alg:coset_sample}
\KwIn{Integers $n,k$ with $0<k\le n$.}{} 

\textbf{Sample $U_1,U_2 \in B$} by drawing each strict-upper entries i.i.d.\ uniform in $\mathbb{Z}_2$.\\

\textbf{Sample $\pi\in S_n$} with Mallows weight $2^{\text{inv}(\pi)}$:
\begin{itemize}
  \item For $i=n,n-1,\dots,1$: draw a uniform integer $c\in\{1,\dots,2^{\,i}-1\}$; set $L_i \gets \lfloor \lg c\rfloor$ so that $\Pr[L_i=t]=2^t/(2^{\,i}-1)$.
  \item Convert the digits $(L_n,\dots,L_1)$ to the permutation $S_{\pi}$.
\end{itemize}
Set $P_y \gets U_1S_{\pi}U_2$. 

\textbf{Return} A basis matrix $A_y \gets P_y\begin{bmatrix} I_{n-r}\\0\end{bmatrix}\in \mathbb{Z}_2^{\,n\times n-r}$ for linear subspace $\operatorname{Im}(A_y)$.
\end{algorithm}

It is important to note that sampling the permutation $\pi$ uniformly from $S_n$ does not yield a uniformly distributed matrix $P_y$. Because the size of the Bruhat cell $B \pi B$ over $\mathbb{Z}_2$ is proportional to $2^{\text{inv}(\pi)}$, where $\text{inv}(\pi)$ is the number of inversions in $\pi$, the permutation must be sampled according to the Mallows distribution \cite{Mallows1,Mallows2} to account for this exponential difference in cell sizes. The sequence of digits $(L_n, \dots, L_1)$ in Algorithm~\ref{alg:coset_sample} represents the Lehmer code of the permutation $\pi$. In a quantum circuit, this can be implemented using CSWAPs with the control on the qubits that encode the digits. While in theory some of the bits of the upper triangular matrix $U_1$ needs to be set to $0$ depending on the chosen permutation, in practice both the upper triangular matrices can be uniformly sampled, and in fact, the random bits generated by the PPRF can be directly used to fill in the non-zero elements of $U_1$ and $U_2$. Binary multiplication by the upper triangular matrices can be implemented on the go by using Toffolis with the control on the qubits encoding the random bits. The translation by the vector $b_y$ can be implemented by CNOTs.

We require $\Omega(n(3n+1)/2 - 1)$ pseudorandom bits to implement the random affine transform evaluation.
A stream cipher mode using the GGM leaf key provides the required pseudorandom bits.

\subsubsection{Generalization to multi-bit messages}

This key generation module can easily be generalized to handle signatures of multi-bit messages \cite{long-message-one-shot,SZ25b}. To support an $l$-bit message, the overall dimension of the secret key state is expanded from $n$ to $n+l$. The algebraic structure is specifically designed so that the first $l$ bits of the generated string will embed the message.

To achieve this, the original full-column rank matrix $A_y \in \mathbb{Z}_2^{n \times (n-r)}$ is augmented into a larger block matrix $\tilde{A}_y \in \mathbb{Z}_2^{(n+l) \times (l + n - r)}$, defined as:

\begin{align}
  \tilde{A}_y = \begin{bmatrix} I_l & 0 \\ B_y & A_y \end{bmatrix},  
\end{align}
where $I_l$ is the $l \times l$ identity matrix, and $B_y \in \mathbb{Z}_2^{n \times l}$ is a rectangular matrix filled with random bits generated via the PPRF alongside the random bits required to implement $A_y$. Similarly, the random translation vector $b_y$ is expanded from $n$ bits to an $(n+l)$-bit string. We can partition this expanded vector into two distinct components: $b_y = b_y^{(1)} \,||\, b_y^{(2)}$, where $b_y^{(1)} \in \mathbb{Z}_2^l$ and $b_y^{(2)} \in \mathbb{Z}_2^n$.

When the augmented affine subspace is evaluated on an input vector constructed by concatenating an $l$-bit string $\omega$ and an $(n-r)$-bit string $z$ (represented as the column vector $[\omega, z]^T$), the transformation yields:
\begin{align}
    \tilde{A}_y \begin{bmatrix} \omega \\ z \end{bmatrix} + \begin{bmatrix} b_y^{(1)} \\ b_y^{(2)} \end{bmatrix}
        =   \begin{bmatrix} I_l & 0 \\ B_y & A_y \end{bmatrix} \begin{bmatrix} \omega \\ z \end{bmatrix}
          + \begin{bmatrix} b_y^{(1)} \\ b_y^{(2)} \end{bmatrix}
        = \begin{bmatrix} \omega + b_y^{(1)} \\ B_y \omega + A_y z + b_y^{(2)} \end{bmatrix}
\end{align}
This outputs an $(n+l)$-bit string where the first $l$ bits cleanly encode the input $\omega$ shifted by $b_y^{(1)}$, and the remaining $n$ bits cryptographically mix $\omega$ and $z$ through the matrices $B_y$ and $A_y$, shifted by $b_y^{(2)}$.

At the end of the KeyGen algorithm, the result is a quantum state containing a superposition of all possible public keys entangled with their corresponding secret key states. Measuring the public key register projects the secret key register into a superposition of all valid signatures associated with that particular public key. The user then holds on to this secret key state until a message needs to be signed. Once a message $m$ is chosen, we move on to the KeySign algorithm.

\subsection{Sign}
We apply a sequence of single qubit measurements followed by adaptive phase flips to target strings in the secret key,
whose first $l$ bits match the message and then measure the remaining $k$ qubits to produce an $l + n$ bits signature.
We first describe the single-bit signing algorithm and then show how to generalize it to $l$-bit signing. The single bit signing algorithm can be constructed as a special case of multi-bit signing with $l=1$. But here we follow the setup in \cite{sz25} where the single bit message is encoded as the first bit of an $n$-bit signature.

After the key-generation subroutine, the signer obtains the public key $y$ together with the secret key quantum state
of the form
\begin{align}
\ket{y}\otimes \ket{sk_y},
\end{align}
where the private state $\ket{sk_y}$ is an equal superposition over an affine coset of a full-rank binary matrix $A_y\in\mathbb{Z}_2^{n\times (n-r)}$:
\begin{align}
\ket{sk_y}
    = \frac{1}{\sqrt{2^{(n-r)}}}
          \sum_{z\in\mathbb{Z}_2^{(n-r)}} \ket{A_y z + b_y}
\end{align}

Here $b_y\in\mathbb{Z}_2^n$ is a pseudorandom bit string which is a part of the KeyGen algorithm. Thus $\ket{sk_y}$ is the uniform superposition over the coset
\begin{align}
C_y := b_y + \operatorname{Im}(A_y),
\qquad
\operatorname{Im}(A_y) = \{ A_y z : z\in\mathbb{Z}_2^{(n-r)}\}.
\end{align}

\subsubsection{Structure of the Coset Space}

We show here that any coset $C=b+\operatorname{Im}(A)$ is either constant on its first coordinate or splits perfectly evenly into strings with first bit~$0$ and first bit~$1$.

For $x = A z + b \in C$, the first bit is
\begin{align}
x_1 = (A z + b)_1
    = A_1 z \;+\; b_1,
\end{align}
where $A_1$ is the first row of $A$. 
Define
\begin{align}
C_0 := \{ x\in C : x_1 = 0\},
\qquad
C_1 := \{ x\in C : x_1 = 1\}.
\end{align}

\paragraph{ Degenerate case: \texorpdfstring{$A_1 = 0$}{A_1=0}.}

If the first row of $A$ is the zero vector, then $A_1 z = 0$ for every $z$,
and therefore
\begin{align}
x_1 = b_1 \quad\text{for all } x\in C.
\end{align}
Which means,
\begin{align}
C_0 = C,\; C_1 = \varnothing
\quad\text{or}\quad
C_1 = C,\; C_0 = \varnothing.
\end{align}
In this case, the coset is constant with respect to its first bit.

\paragraph{Generic case:}
$A_1 \neq 0$.  Then there exists an index $j$ such that $A_{1,j}=1$.
For any $z\in\mathbb{Z}_2^{(n-r)}$, define $z'\in\mathbb{Z}_2^{(n-r)}$ by flipping the
$j$-th bit and keeping all other coordinates fixed:
\begin{align}
z'_j = \lnot z_j, \qquad z'_i = z_i \ \text{for all } i\neq j.
\end{align}
Since $A_{1,j}=1$, we compute
\begin{align}
A_1 z'
   = \sum_{i\neq j} A_{1,i} z_i \;+\; A_{1,j}\,\lnot z_j
   = \sum_{i\neq j} A_{1,i} z_i \;+\; \lnot z_j.
\end{align}
Using $\lnot z_j = z_j + 1$, this becomes
\begin{align}
A_1 z'
   = \sum_{i\neq j} A_{1,i} z_i \;+\; z_j \;+\; 1
   = A_1 z \;+\; 1.
\end{align}
Thus $A_1 z'$ always evaluates to the opposite value of $A_1 z$, showing
that a nonzero linear form on $\mathbb{Z}_2^{(n-r)}$ is necessarily balanced. So this proves that every affine coset $C=b+\operatorname{Im}(A)$ satisfies exactly one of the following:
\begin{align}
\text{Either $x_1$ is constant on $C$, or $C$ splits evenly into $C_0$ and $C_1$.}
\end{align}
In all nondegenerate constructions (where $A_1\neq 0$), the coset is perfectly
balanced.

\subsubsection{Single-Bit Signing}

For a fixed public key $y$, write the corresponding coset
\begin{align}
C_y = b_y + \operatorname{Im}(A_y) \subseteq \mathbb{Z}_2^n.
\end{align}
Assume we are in the generic case where the first row of $A_y$ is nonzero, so
$C_y$ splits evenly according to the first coordinate:
\begin{align}
C_{y,0} := \{ x\in C_y : x_1 = 0 \}, \qquad
C_{y,1} := \{ x\in C_y : x_1 = 1 \},
\end{align}
with
\begin{align}
|C_{y,0}| = |C_{y,1}| = 2^{(n-r)-1}.
\end{align}
We define the two normalized ``half-coset'' states
\begin{align}
\ket{sk_0}
  &:= \frac{1}{\sqrt{2^{(n-r)-1}}} \ket{0} \otimes
     \sum_{x| (0,x)\in C_{y,0}} \ket{x}, \\
\ket{sk_1}
  &:= \frac{1}{\sqrt{2^{(n-r)-1}}} \ket{1} \otimes
     \sum_{x| (1,x)\in C_{y,1}} \ket{x}.
\end{align}
So that the full coset state can be written as
\begin{align}
\ket{sk_y}
  = \frac{1}{\sqrt{2^{(n-r)}}}
    \sum_{x\in C_y} \ket{x}
  = \frac{1}{\sqrt{2}}\bigl( \ket{sk_0} + \ket{sk_1} \bigr).
\end{align}
Here, we write $\ket{sk} := \ket{sk_y}$ in what follows.

Given a one-bit message $m\in\{0,1\}$, the signer uses the first qubit of
$\ket{sk}$ as follows:
\begin{enumerate}
  \item Measure the first qubit of $\ket{sk}$ in the computational basis,
        obtaining an outcome $c\in\{0,1\}$.
        The post-measurement state collapses to $\ket{sk_c}$.

  \item If $c = m$, the signer measures the remaining $n-1$ qubits in the
        computational basis.  The resulting classical string
        $x \in C_{y,m}$ serves as the signature.

   \item If $c \neq m$, apply the logical $X$ operator ($X_L$) to swap the half-coset states, transforming the state to $\ket{sk_m}$. Measure the remaining qubits to obtain the signature $x \in C_{y,m}$.
\end{enumerate}
In all cases, the final classical output lies in the desired half-coset
$C_{y,m}$, i.e.\ its first bit equals the message bit $m$.

\paragraph{Logical $X$ on the Coset Qubit}
To sign a message $m$ that differs from our measured bit $b$, we need a reliable operation that swaps the half-coset states without measuring them.
We construct this logical $X$ operator using a phase oracle in the Fourier domain.

The two orthonormal states $\ket{sk_0}$ and $\ket{sk_1}$ span a
two-dimensional logical qubit.  We define the logical $X$ operator on this
encoded qubit by
\begin{align}
X_L
  := \ket{sk_0}\!\bra{sk_1}
   + \ket{sk_1}\!\bra{sk_0}.
\end{align}
By construction,
\begin{align}
X_L \ket{sk_0} = \ket{sk_1}, \qquad
X_L \ket{sk_1} = \ket{sk_0}.
\end{align}

It is convenient to introduce the symmetric and antisymmetric combinations
\begin{align}
\ket{sk}
  := \frac{1}{\sqrt{2}}\bigl( \ket{sk_0} + \ket{sk_1} \bigr), \qquad
\ket{sk'}
  := \frac{1}{\sqrt{2}}\bigl( \ket{sk_0} - \ket{sk_1} \bigr),
\end{align}
which form an orthonormal basis of the same logical subspace.  We now compute
the action of $X_L$ on these states:
\begin{align*}
X_L \ket{sk}
  &= \frac{1}{\sqrt{2}}
     \bigl( X_L \ket{sk_0} + X_L \ket{sk_1} \bigr)
   = \frac{1}{\sqrt{2}}
     \bigl( \ket{sk_1} + \ket{sk_0} \bigr)
   = \ket{sk},\\[4pt]
X_L \ket{sk'}
  &= \frac{1}{\sqrt{2}}
     \bigl( X_L \ket{sk_0} - X_L \ket{sk_1} \bigr)
   = \frac{1}{\sqrt{2}}
     \bigl( \ket{sk_1} - \ket{sk_0} \bigr)
   = -\,\ket{sk'}.
\end{align*}
Thus $\ket{sk}$ and $\ket{sk'}$ are eigenstates of $X_L$ with eigenvalues
$+1$ and $-1$, respectively:
\begin{align}
X_L \ket{sk} = +\,\ket{sk},
\qquad
X_L \ket{sk'} = -\,\ket{sk'}.
\end{align}
In this basis $\{\ket{sk},\ket{sk'}\}$, the logical $X_L$ plays the role of
a Pauli-$Z$ operator, while in the basis $\{\ket{sk_0},\ket{sk_1}\}$ it acts
as a standard Pauli-$X$ that swaps the two half-coset states.

\paragraph{Logical $X$ from Hadamards and a Phase on $\operatorname{Im}(A_y)^\perp$}

The two states
\begin{align}
\ket{sk}=\frac{\ket{sk_0}+\ket{sk_1}}{\sqrt{2}},\qquad
\ket{sk'}=\frac{\ket{sk_0}-\ket{sk_1}}{\sqrt{2}}
\end{align}
form an orthonormal basis for the two-dimensional coset subspace associated
with $C_y$.  Their structure becomes especially transparent after applying a
global Hadamard transform. Let us look at how a global Hadamard gate will act on these states. For any matrix $A_y$ of full column rank,
\begin{align}
H^{\otimes n} \ket{sk} &= \frac{1}{\sqrt{2^{n-r}}}\sum_{z \in \{0,1\}^{n-r}} H^{\otimes n} \ket{A_y z + b_y}  \\
&= \frac{1}{\sqrt{2^{n-r}}}\sum_{z \in \{0,1\}^{n-r}} \frac{1}{\sqrt{2^n}}\sum_{\omega\in \{0,1\}^{n}} (-1)^{\omega . (A_y z + b_y)} \ket{\omega} \\
&= \frac{1}{\sqrt{2^{n-r}}} \frac{1}{\sqrt{2^n}} \sum_{\omega\in \{0,1\}^{n}} (-1)^{\omega.b_y} \left(\sum_{z \in \{0,1\}^{n-r}} (-1)^{(A_y^{T}\omega).z}\right) \ket{\omega} \\
   &= \frac{1}{\sqrt{2^r}}
     \sum_{\omega\in\operatorname{Im}(A_y)^\perp}
     (-1)^{\omega\cdot b_y}\,\ket{\omega},
\end{align}
where
\begin{align}
\operatorname{Im}(A_y)^\perp := \{\omega\in\mathbb{Z}_2^n : A_y^T\omega = 0\}
\end{align}
is the orthogonal complement of the column space of $A_y$.
Thus $H^{\otimes n}\ket{sk}$ has support exactly on
$\operatorname{Im}(A_y)^\perp$.

We now analyse the Hadamard action on the other state,
\begin{align}
\ket{sk'} := \frac{1}{\sqrt{2}}\left(\ket{sk_0} - \ket{sk_1}\right).
\end{align}
Note that $\ket{sk'}$ is a uniform superposition over the full coset $C_y = b_y + \operatorname{Im}(A_y)$, but with a relative phase of $-1$ applied to elements where the first bit is $1$. Let $e_1 = (1,0, \dots, 0)$ be the unit vector bit string corresponding to the first coordinate. For any vector $x = A_y z + b_y \in C_y$, its first bit is exactly the dot product $e_1 \cdot (A_y z + b_y)$. Thus, we can write $\ket{sk'}$ algebraically as:
\begin{align}
\ket{sk'} = \frac{1}{\sqrt{2^{n-r}}} \sum_{z \in \{0,1\}^{n-r}} (-1)^{e_1 \cdot (A_y z + b_y)} \ket{A_y z + b_y}.
\end{align}

Applying the global Hadamard transform $H^{\otimes n}$ gives:
\begin{align}
H^{\otimes n} \ket{sk'}
  &= \frac{1}{\sqrt{2^{n-r}}} \sum_{z \in \{0,1\}^{n-r}} (-1)^{e_1 \cdot (A_y z + b_y)} H^{\otimes n} \ket{A_y z + b_y} \\
  &= \frac{1}{\sqrt{2^{n-r}}} \frac{1}{\sqrt{2^n}} \sum_{z \in \{0,1\}^{n-r}} \sum_{\omega \in \{0,1\}^n} (-1)^{e_1 \cdot (A_y z + b_y)} (-1)^{\omega \cdot (A_y z + b_y)} \ket{\omega}.
\end{align}

Since we are working on $\mathbb{Z}_2$, we can combine the phases:
\begin{align}
H^{\otimes n} \ket{sk'}
  &= \frac{1}{\sqrt{2^{n-r}}} \frac{1}{\sqrt{2^n}} \sum_{\omega \in \{0,1\}^n} \sum_{z \in \{0,1\}^{n-r}} (-1)^{(\omega + e_1) \cdot (A_y z + b_y)} \ket{\omega} \\
  &= \frac{1}{\sqrt{2^{n-r}}} \frac{1}{\sqrt{2^n}} \sum_{\omega \in \{0,1\}^n} (-1)^{(\omega + e_1) \cdot b_y} \left( \sum_{z \in \{0,1\}^{n-r}} (-1)^{(A_y^T (\omega + e_1)) \cdot z} \right) \ket{\omega}.
\end{align}

The inner sum over $z$ evaluates to $2^{n-r}$ if $A_y^T (\omega + e_1) = 0$, and $0$ otherwise. This condition is equivalent to requiring that $\omega + e_1 \in \operatorname{Im}(A_y)^\perp$. Shifting $e_1$ to the other side, we define the support slice $S$ as:
\begin{align}
S := \operatorname{Im}(A_y)^\perp + e_1 = \{ \omega \in \mathbb{Z}_2^n : \omega + e_1 \in \operatorname{Im}(A_y)^\perp \}.
\end{align}
Thus, the state simplifies to:
\begin{align}
H^{\otimes n} \ket{sk'} = \frac{1}{\sqrt{2^r}} \sum_{\omega \in S} (-1)^{(\omega + e_1) \cdot b_y} \ket{\omega}.
\end{align}

Because we assumed the generic case where the first row of $A_y$ is nonzero, the unit vector $e_1$ is not orthogonal to the column space of $A_y$, meaning $e_1 \notin \operatorname{Im}(A_y)^\perp$. Consequently, the shifted subspace $S$ and the original subspace $\operatorname{Im}(A_y)^\perp$ are completely disjoint:
\begin{align}
\operatorname{Im}(A_y)^\perp \cap S = \varnothing.
\end{align}

Hence the Fourier supports of $\ket{sk}$ and $\ket{sk'}$ are completely disjoint. Now we can define a diagonal operator such that
\begin{align}
D = \sum_{\omega\in\mathbb{Z}_2^n}
       \phi(\omega)\,\ket{\omega}\!\langle\omega|,
\qquad
\phi(\omega)=
\begin{cases}
1,& \text{if }A_y^T\omega=0,\\[4pt]
-1,& \text{otherwise}.
\end{cases}
\end{align}
$D$ can also be written as following,

\begin{align}
D = e^{i\pi \sum\limits_{\omega\notin \operatorname{Im}(A_y)^\perp} \ket{\omega}\bra{\omega}}
\end{align}

Combining all these we get 

\begin{align}
X_L = H^{\otimes n} D H^{\otimes n}
\end{align}
\subsubsection{Multi-bit Signing Using Iterative Measurements}

The one-bit signing algorithm generalizes to $l \sim O(\mathrm{poly}(\lambda))$ bits by expanding the transformation matrix.
We define the augmented binary matrix $\tilde{A}_y \in \mathbb{Z}_2^{(n+l) \times (n+l-r)}$ as follows:

\begin{align}
    \tilde{A}_y = 
    \begin{pmatrix}
    I_l & 0 \\
    B_y & A_y
    \end{pmatrix}
\end{align}

where $B_y \in \mathbb{Z}_2^{n \times l}$ is a matrix of random bits and $A_y \in \mathbb{Z}_2^{n \times (n-r)}$.
This matrix acts on a combined vector $(w \parallel z) \in \mathbb{Z}_2^{l + (n-r)}$.
Given a random shift vector $b_y = (b_y^{(l)} \parallel b_y^{(n)}) \in \mathbb{Z}_2^{n+l}$, the state preparation follows the affine transformation:

\begin{align}
    \ket{\psi} = \frac{1}{\sqrt{2^{n+l-r}}} \sum_{w \in \mathbb{Z}_2^l} \sum_{z \in \mathbb{Z}_2^{n-r}} \ket{w + b_y^{(l)}} \ket{B_y w + A_y z + b_y^{(n)}}
\end{align}

\paragraph{The Iterative Measurement Procedure}

To sign an $l$-bit message $m \in \mathbb{Z}_2^l$, the signer measures the first $l$ qubits sequentially.
Let $p_i$ denote the measurement outcome of the $i$-th qubit.

\begin{itemize}
    \item \textbf{Comparison:} If $p_i = m_i$, the signer proceeds to measure the $(i+1)$-th qubit.
    \item \textbf{Correction:} If $p_i \neq m_i$, the state has collapsed into the ``wrong'' subspace for the $i$-th bit. To recover, the signer applies the bit-specific coset subspace flip gate $X_{L,i}$ before proceeding to the next bit.
\end{itemize}

The flip gate is defined as:
\begin{align} \label{eq:flip}
    X_{L,i} = H^{\otimes (n+l)} D_i H^{\otimes (n+l)}
\end{align}

As an example let's discuss what happens when we measure the first qubit and get the wrong measurement, $p_1 \neq m_1$. 

The measurement projects the state such that the first bit is fixed to $p_1$. Since the first register is $\ket{w + b_y^{(l)}}$, this restricts the summation to values of $w$ where $w_1 + (b_y^{(l)})_1 = p_1$. Let $w_{>1}$ denote the remaining $l-1$ free bits of $w$. The post-measurement state becomes:

\begin{align}
    \ket{\psi_{\text{bad}}} = \frac{1}{\sqrt{2^{n+l-r-1}}} \sum_{\substack{w \in \mathbb{Z}_2^l \\ w_1 = p_1 + (b_y^{(l)})_1}} \sum_{z \in \mathbb{Z}_2^{n-r}} \ket{w + b_y^{(l)}} \ket{B_y w + A_y z + b_y^{(n)}}
\end{align}

The signer applies $H^{\otimes (n+l)}$ to move to the dual space, introducing dual variables $u^{(l)} \in \mathbb{Z}_2^l$ and $u^{(n)} \in \mathbb{Z}_2^n$:

\begin{align}
    H^{\otimes (n+l)} \ket{\psi_{\text{bad}}} = \frac{1}{\sqrt{2^{n+l-r-1}} \sqrt{2^{n+l}}} \sum_{w_1 =p_1} \sum_{w_{>1}} \sum_{z} \sum_{u^{(l)}, u^{(n)}} (-1)^{\Phi} \ket{u^{(l)}} \ket{u^{(n)}}
\end{align}

where the phase $\Phi$ is given by the inner products:

\begin{align}
    \Phi = u^{(l)} \cdot (w + b_y^{(l)}) + u^{(n)} \cdot (B_y w + A_y z + b_y^{(n)})
\end{align}

We rearrange the phase to group the terms by the summation variables $w$ and $z$:

\begin{align}
    \Phi &= \underbrace{(A_y^T u^{(n)}) \cdot z}_{\text{forces } A_y^T u^{(n)} = 0} + \underbrace{(u^{(l)} + B_y^T u^{(n)}) \cdot w}_{\text{splits into } w_1 \text{ and } w_{>1}} \nonumber \\
    &\quad + \underbrace{u^{(l)} \cdot b_y^{(l)} + u^{(n)} \cdot b_y^{(n)}}_{\text{constant phase}}
\end{align}

The sums over $z$ and the free variables $w_{>1}$ act as interference filters:
\begin{itemize}
    \item Summing over $z \in \mathbb{Z}_2^{n-r}$ evaluates to $0$ unless $A_y^T u^{(n)} = 0$.
    \item Summing over the $l-1$ free bits of $w_{>1}$ evaluates to $0$ unless $(u^{(l)} + B_y^T u^{(n)})_j = 0$ for all $j \in \{2, \dots, l\}$.
\end{itemize}

Crucially, because $w_1$ was fixed by the measurement, there is no sum over $w_1$. Therefore, the first component of the message constraint is relaxed.

The dual state has support only where those constraints are met. For the unconstrained bit, let $v_1 = u^{(l)}_1 + (B_y^T u^{(n)})_1 \in \{0, 1\}$. The phase contribution from the fixed $w_1$ bit becomes $(-1)^{v_1(p_1 + (b_y^{(l)})_1)}$. The state simplifies to:

\begin{align}
    H^{\otimes (n+l)} \ket{\psi_{\text{bad}}} \propto \sum_{\substack{u^{(l)}, u^{(n)} \\ A_y^T u^{(n)} = 0 \\ \forall j \neq 1, \; u^{(l)}_j = (B_y^T u^{(n)})_j}} (-1)^{v_1(p_1 + (b_y^{(l)})_1) + \dots} \ket{u^{(l)}} \ket{u^{(n)}}
\end{align}

Because $v_1 = u^{(l)}_1 + (B_y^T u^{(n)})_1$ is free to take values $0$ or $1$, the dual state splits evenly across these two values. Applying the oracle $D_1$ with the phase $(-1)^{u^{(l)}_1 + (B_y^T u^{(n)})_1}$ directly toggles the sign between these two halves, which perfectly corresponds to swapping the primal state from $p_1$ to $\lnot p_1$ (the correct message bit $m_1$) once the Hadamards are applied again.

\paragraph{The Dual-Space Phase Oracle $D_i$}

The operator $D_i$ in Eq. \ref{eq:flip} is defined in the Hadamard (dual) basis. Let $u = (u^{(l)} \parallel u^{(n)})$ be the dual variables. Before any measurement, the state satisfies the constraint $u^{(l)}_j + (B_y^T u^{(n)})_j = 0$ for all $j \in \{1, \dots, l\}$. 

If the measurement of the $i$-th qubit yields an incorrect outcome $p_i \neq m_i$, the state collapses and the $i$-th primal variable is fixed. This relaxes the corresponding $i$-th dual constraint. Consequently, the term $u^{(l)}_i + (B_y^T u^{(n)})_i$ takes values $0$ and $1$ with equal probability across the superposition.

We define the phase oracle $D_i$ to act on this bit:
\begin{align}
    D_i \ket{u^{(l)}} \ket{u^{(n)}} = (-1)^{u^{(l)}_i + (B_y^T u^{(n)})_i} \ket{u^{(l)}} \ket{u^{(n)}}
\end{align}

Applying $D_i$ introduces a $-1$ phase to the subspace where $u^{(l)}_i + (B_y^T u^{(n)})_i = 1$. After applying the subsequent Hadamard transforms to return to the primal space, this phase flip corrects the $i$-th bit from $p_i$ to $m_i$. The state of all other coordinates $j \neq i$ remains unchanged.

\paragraph{Signature Extraction}

Because the localized flip operators $X_{L,i} = H^{\otimes (n+l-i)} D_i H^{\otimes (n+l-i)}$ commute with the underlying subspace constraints and act as the identity on the logical subspaces of $j \neq i$, the iterative procedure successfully forces the first $l$ qubits to match the message $m$ bit-by-bit without disturbing previously measured bits. 

Once all $l$ bits of the first register match the message $m \in \mathbb{Z}_2^l$, the first register has collapsed entirely to the classical string $m$. From the perspective of the initial state preparation, this corresponds to fixing the internal variable $w$ such that $w + b_y^{(l)} = m$, which implies $w = m + b_y^{(l)}$. 

Substituting this fixed $w$ into the second register, the state of the remaining $n$ qubits is a pure superposition over the remaining free variables $z \in \mathbb{Z}_2^{n-r}$:

\begin{align}
    \ket{\psi_{\text{final}}^{(n)}} = \frac{1}{\sqrt{2^{n-r}}} \sum_{z \in \mathbb{Z}_2^{n-r}} \ket{B_y (m + b_y^{(l)}) + A_y z + b_y^{(n)}}
\end{align}

The signer then measures these remaining $n$ qubits in the computational basis. The resulting bitstring $\sigma \in \mathbb{Z}_2^n$ must take the form $\sigma = B_y (m + b_y^{(l)}) + A_y z + b_y^{(n)}$ for some $z$. Rearranging this equation yields the classical verification condition:

\begin{align}
    \sigma + b_y^{(n)} + B_y(m + b_y^{(l)}) = A_y z \in \operatorname{Im}(A_y)
\end{align}

This confirms that the measured signature $\sigma$ is valid and consistent with both the message $m$ and the public key $y$.

\subsubsection{Multi-bit Signing Using Global Measurement and Translation}

To sign an $l$-bit message $m \in \mathbb{Z}_2^l$ using a global approach, the signer performs a simultaneous computational basis measurement of the first $l$ qubits of the state $\ket{\psi}$. 

\paragraph{Global Correction Procedure}
Let $p\in \mathbb{Z}_2^l$ denote the observed $l$-bit measurement outcome. This measurement projects the first register into the state $\ket{p}$, collapsing the remaining unmeasured qubits into the corresponding valid signature subspace for $p$. If $p \neq m$, the signer applies a global translation operator $X_L$ to the entire $(n+l)$ qubit post-measurement state to shift the state into the target subspace associated with the message $m$.

The global translation gate is defined as:
\begin{align}
    X_L = H^{\otimes (n+l)} D_{p \to m} H^{\otimes (n+l)}
\end{align}

\paragraph{Global Dual Phase Oracle}
The oracle $D_{p \to m}$ leverages the Fourier shift theorem to account for the displacement across all $l$ message dimensions simultaneously. Let $d = p + m$ be the $l$-bit displacement vector. In the dual basis $(u^{(l)},u^{(n)})$, the oracle applies the phase shift only on the valid dual subspace satisfying $A_y^T u^{(n)}=0$:

\begin{align}
D_{p \to m}\ket{u^{(l)}}\ket{u^{(n)}}
=
\begin{cases}
(-1)^{(u^{(l)}+B_y^T u^{(n)})\cdot d}
\ket{u^{(l)}}\ket{u^{(n)}},
& A_y^T u^{(n)}=0,\\[4pt]
\ket{u^{(l)}}\ket{u^{(n)}},
& A_y^T u^{(n)}\neq 0.
\end{cases}
\end{align}

For states arising from the legitimate signing procedure, the first Hadamard transform already restricts the dual-space support to $A_y^T u^{(n)}=0$. Therefore, on the honest state the oracle acts exactly as before. Expanding the phase term on this valid subspace yields $u^{(l)}\cdot d+u^{(n)}\cdot B_y d$. Upon applying the subsequent global Hadamard transforms, this phase gradient translates the primal state of the first register by $d$ (mapping $p$ to $m$) and the entangled second register by $B_y d$, satisfying the affine structure of the augmented matrix $\tilde{A}_y$.

The condition $A_y^T u^{(n)}=0$ is not required for the correctness of the honest signing procedure, since this condition is already satisfied if the implementation is error free. We include it explicitly in the definition of $D_{p\to m}$ as an additional safeguard. In particular, without this restriction the same translation operation can act on an arbitrary computational-basis input, allowing a previously obtained valid pair $(m_0,\sigma_0)$ to be reintroduced into the signing circuit and translated toward a signature for a different message $m_1$. Restricting the phase operation to the valid dual subspace prevents $D_{p\to m}$ from acting as an unrestricted translation on such inputs and is introduced specifically to block this simple re-signing attack.

To verify the global translation, we trace the state evolution through the application of $X_L$. 

\textbf{1. Post-Measurement State:} 
When the first register is measured as $p \in \mathbb{Z}_2^l$, the internal variable $w$ collapses such that $w + b_y^{(l)} = p$, which implies $w = p + b_y^{(l)}$. The normalized post-measurement state becomes a superposition over the free variables $z \in \mathbb{Z}_2^{n-r}$:
\begin{align}
    \ket{\psi_{\text{post}}} = \frac{1}{\sqrt{2^{n-r}}} \sum_{z \in \mathbb{Z}_2^{n-r}} \ket{p} \ket{B_y (p + b_y^{(l)}) + A_y z + b_y^{(n)}}
\end{align}

\textbf{2. First Hadamard Transform:} 
Applying $H^{\otimes (n+l)}$ maps the state to the dual space. The summation over $z$ acts as an interference filter, restricting the support to dual variables where $A_y^T u^{(n)} = 0$:
\begin{align}
    \ket{\psi_{\text{dual}}} \propto \sum_{\substack{u^{(l)} \in \mathbb{Z}_2^l, \; u^{(n)} \in \mathbb{Z}_2^n \\ A_y^T u^{(n)} = 0}} (-1)^{\Phi_{\text{old}}} \ket{u^{(l)}} \ket{u^{(n)}}
\end{align}
where the initial dual phase is:
\begin{align}
    \Phi_{\text{old}} = u^{(l)} \cdot p + u^{(n)} \cdot \bigl(B_y (p + b_y^{(l)}) + b_y^{(n)}\bigr)
\end{align}

\textbf{3. Applying the Oracle $D_{p \to m}$:} 
Since $\ket{\psi_{\text{dual}}}$ has support only on dual vectors satisfying $A_y^T u^{(n)} = 0$, the oracle applies the phase shift $(-1)^{(u^{(l)} + B_y^T u^{(n)}) \cdot d}$ to every component in the support, where $d = p + m$. Using the property $(B_y^T u^{(n)}) \cdot d = u^{(n)} \cdot B_y d$, the new phase $\Phi_{\text{new}}$ becomes:
\begin{align}
    \Phi_{\text{new}} &= \Phi_{\text{old}} + u^{(l)} \cdot d + u^{(n)} \cdot B_y d \nonumber \\
    &= u^{(l)} \cdot (p + d) + u^{(n)} \cdot \bigl(B_y (p + d) + B_y b_y^{(l)} + b_y^{(n)}\bigr)
\end{align}
Since $d = p + m$, we have $p + d = m$. Substituting this yields:
\begin{align}
    \Phi_{\text{new}} = u^{(l)} \cdot m + u^{(n)} \cdot \bigl(B_y (m + b_y^{(l)}) + b_y^{(n)}\bigr)
\end{align}

\textbf{4. Second Hadamard Transform:} 
The phase $\Phi_{\text{new}}$ perfectly matches the dual-space representation of a state localized at $m$ in the first register and affine-shifted by $B_y(m + b_y^{(l)}) + b_y^{(n)}$ in the second register. Applying the final $H^{\otimes (n+l)}$ returns the state to the primal basis:
\begin{align}
    X_L \ket{\psi_{\text{post}}} = \frac{1}{\sqrt{2^{n-r}}} \sum_{z \in \mathbb{Z}_2^{n-r}} \ket{m} \ket{B_y (m + b_y^{(l)}) + A_y z + b_y^{(n)}}
\end{align}

This proves that the first register has been perfectly shifted to the target message $m$, while the second register remains in a uniform superposition over the valid signature subspace corresponding exactly to $m$.

\paragraph{Complexity and Success}
This global approach reduces quantum circuit depth compared to iterative bit-flipping, requiring only two layers of Hadamard transforms separated by a single phase-kickback oracle. Because the global shift $X_L$ acts as a bijective translation, it preserves the uniform superposition, mapping the entire set of valid signatures for $p$ directly onto the set of valid signatures for $m$. The signer then measures the remaining $n$ qubits in the computational basis to extract the final signature $\sigma$, completing the protocol.

\subsection{\textsc{Verify}}

As in the original OSS protocol, the verification algorithm in our construction is also a classical, deterministic procedure that checks the validity of the signature $\sigma$ against a message $m$ and a public key $y$. This is illustrated in Figure~\ref{fig:OSS_workflow}. The verification workflow proceeds through a sequence of classical algorithmic steps, leveraging the parameters from the CRS:

\begin{enumerate}
    \item \textbf{GGM Tree Path Traversal:} Using the public key $y$ and an obfuscated verification program containing the hidden key $\klin$, the verifier performs a classical GGM tree search to resolve the corresponding leaf key.
    \item \textbf{Graff Inversion ($\tilde{\textbf{Graff}}$):} Using the derived leaf key, the verifier reconstructs the necessary coset parameters ($P^{-1}_y$, $B_y$, $b_y$). The verifier then computes the pre-image coordinate vector $z$ using the message $m$ and the signature component $\tau$ via the following matrix computation:
    $$z = P_y^{-1} \cdot \left( B_y \cdot\left( m+b_y^{(l)}\right) + \tau + b_y^{(n)} \right)$$
    The inverse matrix $P^{-1}_y$ can be efficiently applied by taking the sequence of gates used in the \textsc{Graff} circuit and inverting it.
    \item \textbf{Validity Constraints:} The signature is universally accepted if and only if the outputs of the following two independent comparators both yield true (i.e., evaluated via a logical AND gate):
    \begin{itemize}
        \item \textit{Zero-Padding Check:} The verifier ensures that the last $r$ bits of the computed vector $z$ are all zeros, validating that $z_{n-r \dots n} \overset{?}{=} 0^{\otimes r}$.
        \item \textit{Message Bit Check:} The verifier ensures that the first $l$ bits of the signature $\sigma$ exactly match the signed message $m$, confirming that $\sigma_{1 \dots l} \overset{?}{=} m$.
    \end{itemize}
\end{enumerate}

\section{Logical resources}
\label{sec:resources}

\subsection{Logical qubits}
\textsc{KeyGen} and \textsc{Sign} both require enough room for
 a public key,
 a signature,
 a \textsc{GGM.Lookup},
 and a \textsc{Graff}.
In symbols that is $\#Q_{\textsc{OSS}} = r + n + \#Q_\textsc{GGM.Lookup} + \#Q_\textsc{Graff}$;
however, due to the sequential nature of the circuits, there are some gestalt savings available.

We calculated above that \textsc{Graff} requires $\text{bk}+n+1$ qubits.
We also know that \textsc{GGM.Lookup} keeps checkpoints in only $\lceil-\lg(1 - 2^{-R}(r+1))\rceil$
of its $R := \lceil \lg(r+2) \rceil$ key registers.
This leaves $f = R - \lceil-\lg(1 - 2^{-R}(r+1))\rceil$ free key registers
that can be repurposed to satisfy the \textsc{Graff} requirements.
Moreover, \textsc{GGM.Lookup} adds an additional $2\text{bk}$ qubits when run with small keys.
Let $Q_{\text{add}}$ denote the extra qubit overhead required when these repurposed free registers are insufficient to cover the \textsc{Graff} requirements, defined as: $
    \varepsilon_{\text{add}} = \max(\text{bk}+n+1 - f - \varepsilon_{\text{small keys}}, 0)$.

Tying this all together gives the total logical qubit count:
\begin{equation} \label{eq:logical_qubits}
    \#Q_\textsc{OSS} = r + n + R\kappa + \varepsilon_{\text{add}}
\end{equation}

The first nine configurations are evaluated and tabulated in Table~\ref{tab:QubitCount1bitOSS}.
\begin{table}[h]
\caption{Logical qubit counts for single-bit-message OSS.}
\label{tab:QubitCount1bitOSS}
\begin{tabular}{r|S|S|S|S|S|S|S|S|S}
\hline
    \hline
$\lambda$ &   2 &   3 &   4 &   5 &   6 &    7 &    8 &    9 &   10 \\
r & 4 & 12 & 24 & 40 & 60 & 84 & 112 & 144  & 180 \\
n & 10 & 21 & 36 & 55 & 78 & 105 & 136 & 171  & 210 \\
$\kappa$& 64 &     64 &      64 &      64 &       96 &      128 &       144 &       192  &      256 \\
\hline
\hline
    $\#Q_\textsc{OSS}$ & 206 & 289 & 380 & 479 & 745 & 1085 & 1256 & 1851 & 2438 \\
    \hline
    \hline
\end{tabular}
\end{table}

%

\subsection{Logical gates}
\subsubsection{Key Generation}
As stated above, \textsc{GGM.Lookup} has \textsc{NextKey} step complexity $\Theta(r^{\lg(3)})$.
Each \textsc{NextKey} circuit makes
\begin{tabular}{ c | c | c}
    X   &  CX & \textsc{Cipher} \\
\hline
    2 & 1 & 1
\end{tabular}
calls.

The \textsc{Graff} gate complexity scales with signature size, and is tabulated in Table~\ref{tab:graff_gc}.
We lower bound the number of PRG calls necessary to step through the pseudorandom stream.
Multiplication by the triangular matrices contributes only to the CCX count.
The permutation instructions are decoded into CSWAPs.
And the coset shift is implemented by CX gates.

\begin{table}[h!]
\caption{The \textsc{Graff} gate complexity.}
\label{tab:graff_gc}
\begin{center}
\begin{tabular}{ r | c }
    X          & $2n(n+1) - 4$           \\
    CX         & $n(n+1) - 2 + n$        \\
    CCX        & $2n^2-2$                \\
    CSWAP      & $(n-1)n(n+1)/6$         \\
    $bk$ \ PRG & $\Omega(n(3n+1)/2 - 1)$ \\
\end{tabular}
\end{center}
\end{table}

The \textsc{KeyGen} circuit factors into
 two \textsc{GGM.Lookup}s,
 one \textsc{Graff},
 $n$ Hadamards,
 and $r$ measurements.
The subcircuit resource estimates are combined, evaluated, and tabulated in Table~\ref{tab:keygen}.
We note that the $\lambda=9$ circuit has a higher total gate count than the
$\lambda=10$ circuit; this is because $\lambda=9$ runs a small key \textsc{GGM.Lookup}.

\begin{table}[h!]
\caption{Gate counts for the \textsc{KeyGen} algorithm}
\label{tab:keygen}
\begin{tabular}{r|S|S|S|S|S|S|S|S|S}
\hline
    \hline
$\lambda$ & 2 & 3 & 4 & 5 & 6 & 7 & 8 & 9 & 10 \\
r & 4 & 12 & 24 & 40 & 60 & 84 & 112 & 144  & 180 \\
n & 10 & 21 & 36 & 55 & 78 & 105 & 136 & 171  & 210 \\
$\kappa$& 64 &     64 &      64 &      64 &       96 &      128 &       144 &       192  &      256 \\
\hline
  H &     10 &     21 &      36 &      55 &       78 &      105 &       136 &       171 &       210 \\
  X &  31084 & 139018 &  438078 & 1106092 &  2640302 &  9415012 &  41258684 & 149478128 & 100206816 \\
 CX & 175576 & 787037 & 2487166 & 6298781 & 14508496 & 50691945 & 131321638 & 468369837 & 521750744 \\
CCX &  19819 &  89460 &  284792 &  727016 &  1665453 &  5638584 &  22985874 &  80847916 &  55904687 \\
\hline\hline
    & 226489 & 1015536 & 3210072 & 8131944 & 18814329 & 65745646 & 195566332 & 698696052 & 677862457 \\
\hline\hline
\end{tabular}
\end{table}

\subsubsection{Signing}
Recall that a measurement on the first qubit of a coset state
collapses the sum into a collection of subterms with the same leading bit.
Applying an \textsc{Equivocate} circuit switches between these collections. The circuit \textsc{Equivocate} is the circuit realization for the global dual phase oracle $D_{p\rightarrow m}$.
And we use \textsc{Gr*} to denote the circuit that applies the transpose of the linear part of $\textsc{Graff}$.
An \textsc{Equivocate} circuit requires,
\begin{align*}
\gatenorm{Equivocate} =   2\gatenorm{GGM.Lookup}
                        + 2\gatenorm{Gr*}
                        + 2n\textsc{H}
                        + 2(n-r)\textsc{X}
                        + 2\textsc{C}_{n-r}\textsc{X}
                        + \textsc{Z}
\end{align*}

The $C_jX$ gate can further be decomposed to $2j-3$ CCX and $2j-6$ X gates and a clean ancilla qubit \cite{Khattar2025riseofconditionally}. The signing routine requires an equivocation half the time,
giving an expected gate count:
\begin{align*}
    \mathbf{E} \gatenorm{Sign} = n\textsc{Measure} + \gatenorm{Equivocate}/2
\end{align*}

\begin{table}[h!]
\centering
\caption{Gate counts for single-bit-message \textsc{Sign} that equivocates.}
\label{tab:sign}
\begin{tabular}{r|S|S|S|S|S|S|S|S|S}
\hline\hline
$\lambda$ & 2 & 3 & 4 & 5 & 6 & 7 & 8 & 9 & 10 \\
    r & 4 & 12 & 24 & 40 & 60 & 84 & 112 & 144 & 180 \\
    n & 10 & 21 & 36 & 55 & 78 & 105 & 136 & 171 & 210 \\
    $\kappa$ & 64 &      64 &      64 &      64 &       96 &      128 &      144 &       192  &       256 \\
    \hline
    H   &    20  &      42 &      72 &      110 &      156 &      210 &       272 &       342 &       420 \\
    X   &  42692 &  183828 &  573966 &  1419262 &  3382370 & 11124478 &  44956900 & 157079426 & 112008600 \\
    CX  & 240516 & 1039388 & 3259712 &  8099068 & 18655968 & 60123560 & 143740044 & 493581148 & 585557868 \\
    CCX &  27368 &  119558 &  379162 &   954374 &  2193948 &  6827966 &  25443198 &  85776062 &  63886288 \\
    Z   &      1 &       1 &       1 &        1 &        1 &        1 &         1 &         1 &         1 \\
    \hline\hline
        & 310597 & 1342817 & 4212913 & 10472815 & 24232443 & 78076215 & 214140415 & 736436979 & 761453177 \\
    \hline\hline
\end{tabular}
\end{table}

\subsection{Logical resources for multi-bit signing}
Generalizing to longer messages requires $2l$ additional qubits:
$l$ to store the signature, and another $l$ ancilla qubits for the global measurement protocol.
We add this to our earlier analysis in Eq.~\ref{eq:logical_qubits} to arrive at,
\begin{equation}
\#Q = n+2l + r+\kappa\lceil\lg(r+2)\rceil
\label{eq:LogicalQubitsMultiBit}
\end{equation}

To calculate the extra gates required for both key generation and signing, we will now go through each step of the protocol.

\begin{itemize}
    \item Hadamards: Since the signature size is $n+l$ instead of $n$, we will need $l$ extra Hadamards for this step. 
    \item \textsc{GGM.Lookup}: This step only depends on the size of public key ($r$) and not on the signature size. So the number of gates remain the same.
    \item \textsc{Graff}: We need $nl$ extra random bits to perform the action of $B_y$. This requires $\lceil \frac{nl}{bk}\rceil$ calls to PRG. To realize extra matrix multiplications in the Bruhat decomposition, we need an extra $nl$ CCX and $l$ CX gates.
    \item \textsc{Sign}: The number of extra gates/calls required scales as follows

    \begin{center}
\begin{tabular}{ r | c }
\#\textsc{GGM.Lookup} & 2 \\
    H          & $2l$ \\
    X          & $l$  \\
    CX         & $2l$ \\
    CCX        & $nl$ \\
    $bk$ \ PRG & $nl$ \\
\end{tabular}
\end{center}
\end{itemize}

\section{Concrete security}
\label{sec:security}

\subsection{Parameter choice and discussion on possible attacks}\label{security}

Let $\lambda \in \mathbb N$ be the statistical security parameter. Recall that the OSS Construction $52$ in Shmueli and Zhandry's work~\cite{sz25} defines the following parameters in terms of $\lambda$: $n,r,k,s,d,\kappa$. The parameters $n,r,k,s$ determine the sizes and dimensions of subspaces and cosets while $\kappa$ determines the security level of the underlying pseudorandom function. These parameters are chosen to be sufficiently large to prevent efficient collision finding in the oracle based construction. In the corresponding plain-model construction, indistinguishability obfuscation (iO) is used to translate oracle-based security into plain-model security. In this setting, the cryptographic security parameter $\kappa$ plays a central role and must be selected carefully. Note that the values of the structural parameters $s,r,n,k$ remain unchanged, regardless of whether obfuscation is used.

One subtle point is that~\cite{sz25} use $\lambda$ to describe the statistical security of their constructions. This parameter may be roughly interpreted as an efficient adversary's success probability of finding collisions for the underlying non-collapsing hash function.

In this work, since the information of the $(\textsc{KeyGen}, \textsc{Sign}, \textsc{Verify})$ algorithms, including the oracles $P$, $P^{-1}$, and $D$, will be published in obfuscated forms, an adversary can use any of these oracles to perform attacks. Here, we focus on the values for $r$ and $\kappa$ and only discuss the best possible attacks by both classical and quantum adversaries:

\begin{itemize}
    \item $r$: Determines the number of possible hash digests (and public keys) $y \in \{0,1\}^r$. 
    \begin{itemize}
        \item A classical adversary can use the verification circuit containing the oracles $P^{-1}$ with the information of the public key $y$ to find two pairs of valid signatures with the same public key, running $P^{-1}(y,u)$ until he finds $x$ such that $P^{-1}(y,u)=(x\|0^r)$. The probability of finding such a valid $u$ becomes high after trying $O(\frac{2^{n}}{2^{n-r}}) = O(2^r)$ inputs.
        \item In the quantum case, the adversary can prepare a superposition of all $n$-bit strings together with the known public key as follows: $\sum_{x\in\{0,1\}^n}\ket{x}\ket{y}$. Using the quantisation of the classical verification circuit, the adversary can run Grover's algorithm to find a valid signature in approximately $O(2^{r/2})$ steps.
    \end{itemize}

    Therefore, $r$ must be scaled sufficiently large with the security parameter $\lambda$ so that both attack complexities, $O(2^r)$ classically and $O(2^{r/2})$ quantumly, are computationally infeasible. If $r$ is too small, an adversary can efficiently run either of these processes to find valid signatures under the same public key $y$. Noting that the messages is the first bit of the signature, by repeating the aforementioned process, the adversary may produce either a collision - two valid signatures on the same message - which breaks the security of the underlying hash function, or multiple valid message-signature pairs on distinct messages, hence violating the one-shot property.
    
    \item  $\kappa$: controls the security of the underlying pseudorandom function. If $\kappa$ is small, a quantum adversary can recover the PRF key $\klin$ using Grover search in $O(2^{\kappa/2})$ oracle queries. The knowledge of $\klin$ allows the adversary to reconstruct the coset description $(A_y, b_y)$ and regenerate the valid coset states, hence creating additional valid signatures on the same message.
\end{itemize}

We now discuss our choice for the cryptographic security parameter $\kappa$. Assume that all cryptographic building blocks are subexponentially secure for a subexponential (in $\lambda$) function $f(\lambda) = 2^{\lambda^{\delta}}$ with $0<\delta<1$. Note that the security of the protocol in the plain model construction relies on a carefully chosen parameter set that guarantees the following: for any PPRF key of size~$\kappa$, and for the total number of cosets $2^r$—which is also the number of possible hash digests~$y$—the Bloating the Dual technique ensures that each coset $C_y$ is embedded into a superspace $T_y$ with $C_y\subseteq T_y \subseteq \mathbb{Z}^k_2$.
Under this construction, the value of~$\kappa$ is derived using the bound given in~\cite[Lemma~26, Lemma~56]{sz25} guarantee that an adversary $\mathcal{A}$ cannot tamper with the obfuscated circuit associated with these superspaces in order to exploit the collisions of the underlying hash function:
\begin{align}
    \frac{2^r \cdot \frac{k^2}{\epsilon}}{f(\kappa)} &\leq \frac{1}{512} \\
    \kappa&\geq (9 + 2\lambda(\lambda - 1) + \lambda^\delta + 2\lg(2\lambda^2+\lambda))^{1/\delta}
    \label{secineq}
\end{align}

Appropriate values of $\kappa$ for varying $\lambda$ can be found in Table~\ref{tab:QubitCount1bitOSS}.

\subsection{Attacks on key-generation inputs}
\label{sec:keygen-input-attacks}

Omitting the PPRP $\Pi$ leaves the state produced by honest
key generation unchanged. However, it opens up attacks
based on other choices of the initial state. The following
attacks apply when the coherent key generation circuit can
be evaluated on inputs chosen by the adversary.

Let $V$ denote the key generation circuit before measurement.
We label the public-key register by $Y$ and the remaining
registers by $\mathrm{rest}$. We write
\[
    V = F(H_Y^{\otimes r}\otimes H_{\mathrm{rest}}),
\]
where $H_{\mathrm{rest}}$ consists of the Hadamard gates
applied to the remaining registers. Setting
$\ket{\chi}=H_{\mathrm{rest}}\ket{0}_{\mathrm{rest}}$,
the affine evaluation gives
\[
    F\bigl(\ket{y}_Y\ket{\chi}_{\mathrm{rest}}\bigr)
    = \ket{y}_Y\ket{sk_y}_{\mathrm{rest}}.
\]
Here and below, workspace registers returned to zero
are omitted.

An adversary who knows the public key $y$ can prepare
$H^{\otimes r}\ket{y}$ in the public-key register and
initialize the other registers to zero. Applying the
key generation circuit gives
\[
    V\bigl(H^{\otimes r}\ket{y}\otimes\ket{0}\bigr)
    = \ket{y}\otimes\ket{sk_y}.
\]
Thus, a second secret-key state can be generated after
the public key has been announced. Wrapping the
obfuscation around the Hadamard layer does not prevent
this attack, since the adversary only needs to choose
the input to the full circuit.

A second attack uses entangled inputs to two evaluations
of the key generation circuit. The adversary prepares
$r$ Bell pairs, placing one half of each pair in each
public-key register, and initializes the remaining
registers to zero:
\[
    \ket{\Psi_{\mathrm{in}}}
    =
    \frac{1}{\sqrt{2^r}}
    \sum_{x\in\{0,1\}^r}
    \bigl(\ket{x}_{Y_A}\ket{0}_{\mathrm{rest}_A}\bigr)
    \otimes
    \bigl(\ket{x}_{Y_B}\ket{0}_{\mathrm{rest}_B}\bigr).
\]
Applying Hadamard gates to both halves leaves the
Bell pairs unchanged:
\[
    (H^{\otimes r}\otimes H^{\otimes r})
    \sum_x\ket{x}\ket{x}
    =
    \sum_y\ket{y}\ket{y}.
\]
The entangled input can therefore be supplied before
the Hadamard layers. Applying the two key generation
circuits produces
\[
    (V\otimes V)\ket{\Psi_{\mathrm{in}}}
    =
    \frac{1}{\sqrt{2^r}}
    \sum_y
    \bigl(\ket{y}\ket{sk_y}\bigr)_A
    \otimes
    \bigl(\ket{y}\ket{sk_y}\bigr)_B.
\]
The logical circuit considered here consists of
Hadamard gates and reversible classical operations,
so $V$ has real matrix entries. Hence the paired
evaluation $V\otimes V^*$ is the same as
$V\otimes V$, where $V^*$ denotes entrywise complex
conjugation.

Measuring the public-key registers gives the same
value $y$ in both registers and leaves two copies
of $\ket{sk_y}$. These copies can be used independently
to sign two distinct messages. Although $y$ is
random, this is sufficient to violate the security
definition, which allows the adversary to output
the public key together with the signatures.

To address these attacks, we consider applying an
additional unitary $U$ before the key generation
circuit, with
\[
    U\ket{0}_{Y,\mathrm{rest}}
    = \ket{0}_{Y,\mathrm{rest}}.
\]
The modified circuit is
\[
    \widehat V=VU,
\]
and honest key generation is unchanged because
\[
    \widehat V\ket{0}_{Y,\mathrm{rest}}
    = V\ket{0}_{Y,\mathrm{rest}}.
\]
The action of $U$ on other inputs must be chosen
to prevent the attacks described above.

The condition $U\ket{0}=\ket{0}$ alone is not
sufficient, even if $U$ acts nontrivially on other
inputs. For example, let
\[
    U=U_Y\otimes U_{\mathrm{rest}},
\]
where both factors are real unitaries that fix
their respective all-zero states. In this case,
\[
    (U\otimes U)\ket{\Psi_{\mathrm{in}}}
    = \ket{\Psi_{\mathrm{in}}},
\]
so the Bell-pair attack remains unchanged. The
additional unitary must therefore alter the
correlations used in the attack.

The combined circuit $VU$ would also need to be
obfuscated so that $U$ is not exposed as a separate
operation that the adversary can simply undo.
A security proof must account for the adversary's
access to the full circuit and to the other public
operations of the scheme. We leave the choice of
such a unitary and the corresponding security
proof as requirements for completing this approach.

\subsection{Obfuscation}\label{sec:obfuscation}
\subsubsection{Classical iO} Classical indistinguishability obfuscation (iO) is best viewed as the viable relaxation of general program obfuscation that survived Barak et al.'s impossibility result for virtual-black-box obfuscation \cite{iO}; Goldwasser and Rothblum \cite{GR07} further showed that, for efficient obfuscators, iO is equivalent to best-possible obfuscation. The first candidate for all circuits was given by Garg et al. \cite{GGHRS+16}, but the early multilinear-map era came with severe polynomial overhead. For a size-$s$ fan-out-1 circuit (i.e., Boolean formula), standard Barrington-style preprocessing already yields a branching program of length $s^{3.64}$, and Ananth et al. \cite{ADYS14} analyze a line of constructions whose obfuscation size was $O(s^{10.92})$ under a quadratic-cost assumption before being improved to $O(s^5)$. To overcome the inherent restrictions and conversion overheads of formula-based obfuscation, subsequent works bootstrapped iO to more general computational models. Bitansky and Vaikuntanathan \cite{BV18} showed how to obtain iO from sufficiently succinct functional encryption with obfuscation size $2|C|+\mathrm{poly}(\lambda,d,n)$ for any circuit $C$, where $\lambda$ is the security parameter, $d$ is circuit depth and $n$ is input length, while Ananth, Jain, and Sahai \cite{AJS17} obtained bounded-input (by $L$) iO with size $2|M|+\mathrm{poly}(\lambda, L)$ for any Turing machine $M$. A major conceptual milestone was then achieved by Jain, Lin, and Sahai \cite{JLS26}, who established iO for all polynomial-size circuits from well-founded assumptions, albeit at the level of general polynomial overhead rather than near-linear efficiency. Along this line, Ragavan et al. \cite{RVV25} simplified the assumption set by removing the need for polynomial-stretch $\mathsf{NC}^0$ pseudorandom generators. 

More recent work has improved efficiency in structured settings instead of merely proving existence. Jain and Jin \cite{JJ22} removed the exponential loss in the input length whenever functional equivalence admits short propositional proofs, Ma, Dai, and Shi \cite{MDS25} further reduced the resulting obfuscation and evaluation complexity to quasi-linear in the circuit size and proof size. Parallely, Diamond iO \cite{diamondiO} simplified lattice-based iO by replacing costly recursive encryption with lightweight matrix operations, maintaining an $O(L)$ overhead with respect to the input length while bounding the overall obfuscation size by standard $\mathrm{poly}(|C|)$. Notably, a recent work by Jain et al. \cite{JJMP25} introduced fully succinct iO, which demonstrates that the size of the obfuscated program grows only with the program's secret part, remaining completely independent of the public description and input size. Consequently, this provides an affordable means to obfuscate the puncturable PRF key $\klin$ within the classical portion of our one-shot signature. Overall, the literature suggests a clear spectrum, i.e., while unrestricted classical iO remains a high-degree-polynomial object, structured variants can already be quasi-linear, or even more efficient in the case that only a small classical secret subprogram needs to be hidden. This makes classical iO a plausible choice in our setting, as the resulting overhead can reasonably be expected to remain tolerable. We will discuss it in detail in Sec.~\ref{subsubsec:hide_k_lin}.

\subsubsection{Quantum iO} Quantum iO is substantially less mature than classical iO. Alagic and Fefferman \cite{AF16} formalized several notions of quantum obfuscation and proved strong impossibility results, showing that generic quantum black-box obfuscation fails once an adversary can obtain multiple obfuscated copies, and statistical indistinguishability obfuscation would imply an unlikely complexity collapse. Alagic et al. \cite{ABDS21} later showed that even quantum virtual-black-box obfuscation of classical circuits is generally impossible under quantum-hard LWE. Positive results therefore began only for highly restricted classes. Broadbent and Kazmi \cite{BK21} obtained a construction whose size is exponential in the number of T gates and hence efficient only for a Clifford+T circuit $C$ with T-count $O(\log |C|)$. 
Subsequent work primarily expanded the class of quantum circuits that can be obfuscated, rather than improving the asymptotic efficiency of the constructions. Bartusek et al. \cite{BKNY23} obtained obfuscation for polynomial-size pseudo-deterministic quantum circuits in the classical-oracle model; Coladangelo and Gunn \cite{CG24} introduced quantum state iO, showed that it yields best-possible copy protection for all programs, but only relative to an efficient quantum oracle; and Bartusek, Brakerski, and Vaikuntanathan \cite{BBV24} constructed quantum state obfuscation for pseudo-deterministic quantum programs in the classical-oracle model, heuristically instantiable from quantum-secure classical iO. Huang and Tang \cite{HT25} then extended the scope to unitary (and approximately unitary) quantum programs with quantum inputs and outputs in the classical-oracle model, and the same authors in a very recent preprint \cite{HT26} further claims the first universal obfuscation for arbitrary quantum circuits with quantum inputs and outputs in the classical-oracle model, with a corollary in the quantumly accessible pseudorandom-oracle model from post-quantum subexponential one-way functions and functional encryption.

\subsubsection{Hiding the GGM master key using fully succinct obfuscation}\label{subsubsec:hide_k_lin}
Jain, Jin, Mathialagan, and Paneth~\cite{JJMP25} introduce a fully
succinct obfuscation notion for Turing machines with split descriptions.
In this syntax, a program is written as $M[\mathsf{pub}]$, where
$\mathsf{pub}$ is a public part that may be large, while $M$ is the
secret part that we wish to protect. The fully succinct guarantee says
that the size of the obfuscated program grows only with the size of the
secret part $M$, and not with the public part $\mathsf{pub}$ or the
input length. This is exactly the setting applicable to our PPRF construction, as
the algorithms for GGM evaluation, puncturing, and the subsequent linear-algebraic post-processing are public, while the value we want to hide is the GGM master key $\klin$.

To this end, we only need the fully succinct EF-iO, or equivalently
the fully succinct pv-iO, guaranteed in~\cite{JJMP25}, rather than general
iO for all equivalent machines. Indeed, the two programs we transition between below are not arbitrary equivalent programs: their
equivalence follows from the PV correctness of GGM puncturing by \cite[Lemma IV.12]{JJMP25}. We now
verify the required EF-equivalence condition.

Let $\mathcal{F}_\mathsf{master}$ be the GGM evaluation algorithm we use to construct our PPRF. Let $\mathsf{Post}(y,k_y,\mathsf{pub})=(A_y,b_y)$ denote the deterministic polynomial-time post-processing algorithm used by our construction. Specifically, $\mathsf{Post}$ takes the point $y$, the GGM
leaf key $k_y:=\mathcal{F}_\mathsf{master}(\klin,y)$, and public parameters $\mathsf{pub}$, and returns the matrix-vector pair $(A_y,b_y)$. We note that all randomness used by the post-processing is derived deterministically from the GGM output $k_y$. Thus $\mathsf{Post}$ is a
deterministic polynomial-time computation and can be formalized in PV.

Fix a puncturing set $S\subseteq \{0,1\}^r$, let $k^S$ be the punctured key, and define the table of punctured values as $\Delta_S:=\{(y,k_y): y\in S, k_y=\mathcal{F}_\mathsf{master}(\klin,y)\}$. We use $\Delta_S[y]$ to denote the table lookup value $k_y$ for $y\in S$.

The set $S$ is treated as part of the public description. The original
PPRF program with the master key is
\begin{align}
  M_{\mathsf{master}}[\mathsf{pub},S](y)
  :=
  \mathsf{Post}(y,\mathcal{F}_\mathsf{master}(\klin,y),\mathsf{pub}).
\end{align}
The corresponding PPRF program with the punctured key is
\begin{align}
  M_{\mathsf{punc}}[\mathsf{pub},S](y)
  :=
  \begin{cases}
  \mathsf{Post}(y,\Delta_S[y],\mathsf{pub}),
  & \text{if } y\in S,\\[1.5mm]
  \mathsf{Post}(y,\mathcal{F}_\mathsf{punc}(k^S,y),\mathsf{pub}),
  & \text{if } y\notin S.
  \end{cases}
\end{align}
Here $\mathcal{F}_\mathsf{punc}$ denotes the punctured GGM evaluation algorithm. 

Now we show that the two programs $M_{\mathsf{master}}[\mathsf{pub},S]$ and $M_{\mathsf{punc}}[\mathsf{pub},S]$ agree on every input $y\in\{0,1\}^r$. The proof is by case analysis. First, we suppose $y\notin S$. By  \cite[Lemma IV.12]{JJMP25}, the standard GGM construction of PPRF has a PV proof of
puncturing correctness. Thus PV proves
\begin{align}
  y\notin S
  \quad\Longrightarrow\quad
  \mathcal{F}_\mathsf{punc}(k^S,y)=\mathcal{F}_\mathsf{master}(\klin,y).
\end{align}
Since $\mathsf{Post}$ is deterministic polynomial-time and formalizable
in PV, PV is closed under substituting equal terms into
$\mathsf{Post}$. Therefore PV also proves
\begin{align}
  y\notin S\quad\Longrightarrow\quad\mathsf{Post}(y,\mathcal{F}_\mathsf{punc}(k^S,y),\mathsf{pub})=\mathsf{Post}(y,\mathcal{F}_\mathsf{master}(\klin,y),\mathsf{pub}).
\end{align}
Hence
\begin{align}
  y\notin S
  \quad\Longrightarrow\quad
  M_{\mathsf{punc}}[\mathsf{pub},S](y)
  =
  M_{\mathsf{master}}[\mathsf{pub},S](y).
\end{align}
Now suppose $y\in S$. By construction of the table $\Delta_S$, we
have
\begin{align}
\Delta_S[y]=\mathcal{F}_\mathsf{master}(\klin,y).
\end{align}
Again, substituting this equality into the deterministic PV-formalizable
function $\mathsf{Post}$, PV proves
\begin{align}
  y\in S\quad\Longrightarrow\quad\mathsf{Post}(y,\Delta_S[y],\mathsf{pub})=\mathsf{Post}(y,\mathcal{F}_\mathsf{master}(\klin,y),\mathsf{pub}).
\end{align}
Therefore
\begin{align}
  y\in S
  \quad\Longrightarrow\quad
  M_{\mathsf{punc}}[\mathsf{pub},S](y)
  =
  M_{\mathsf{master}}[\mathsf{pub},S](y).
\end{align}
Combining the two cases, PV proves for every input $y\in\{0,1\}^r$,
\begin{align}
  M_{\mathsf{punc}}[\mathsf{pub},S](y)
  =
  M_{\mathsf{master}}[\mathsf{pub},S](y).
\end{align}
By the propositional translation theorem for PV (see \cite[Theorem IV.2]{JJMP25}, following Cook~\cite{Cook75} and Cook--Reckhow~\cite{CookReckhow79}), this PV proof yields a uniform polynomial-size EF proof of the statement
\begin{align}
  \forall y\in\{0,1\}^r,\qquad
  M_{\mathsf{punc}}[\mathsf{pub},S](y)
  =
  M_{\mathsf{master}}[\mathsf{pub},S](y).
\end{align}
Hence the two split-description programs are uniformly EF-equivalent.

Finally, the formal definition of fully succinct EF-iO requires the two
programs to have the same description length and the same running time. We satisfy this by standard padding: we pad the shorter secret description with dummy bits, and we pad the faster computation with dummy
steps. This padding does not change the functionality of either program and is itself PV-formalizable. Therefore, the padded versions of
$M_{\mathsf{master}}[\mathsf{pub},S]$ and
$M_{\mathsf{punc}}[\mathsf{pub},S]$ satisfy the same-size, same-runtime, and uniform-EF-equivalence conditions required for fully
succinct EF-iO.

\begin{cor}[Obfuscating without exposing the PPRF master key]
The program $M_{\mathsf{master}}$ contains the original GGM master key $\klin$, whereas $M_{\mathsf{punc}}$ contains only the
punctured key $k^S$ and the table of punctured values $\Delta_S$.
Thus, the obfuscation of the PPRF program with a master key is computationally
indistinguishable from the obfuscation of an equivalent program whose secret description no longer contains $\klin$ by the fully succinct EF-iO in \cite{JJMP25}.
\end{cor}

\subsubsection{Conditional quantum obfuscation component}

We stress that the obfuscation step used for hiding the classical GGM/PPRF master key $\klin$ is purely classical. In particular, the corresponding master-key and punctured-key programs are classical split-description Turing machines, and their equivalence is certified by the PV correctness of GGM puncturing. Hence this part can be handled by the fully succinct EF-iO/pv-iO machinery of \cite{JJMP25}.

The situation is different for the main one-shot signature construction proposed in this work. In the original one-shot signature paradigm, the obfuscated object is classical, for instance a suitable obfuscation of a pseudorandom permutation or a permutable PRP, and classical iO is sufficient for this purpose. In our variant, however, we replace the classical PPRP layer by a quantum procedure involving Hadamard gates. As a result, the object that must be hidden is no longer a purely classical program, but a quantum circuit. Classical iO does not apply to such a circuit-level quantum functionality. To instantiate this part of our construction, one would need a quantum indistinguishability obfuscator for the relevant class of Hadamard-based quantum circuits.

At present, we therefore view this quantum one-shot signature component as a conditional and forward-looking instantiation. Existing quantum obfuscation results provide important evidence that quantum iO is a meaningful theoretical primitive, but they do not yet yield a practical plain-model implementation suitable for our scheme.  Thus, our construction separates the currently implementable classical obfuscation layer from the quantum obfuscation layer that remains conditional on future progress in efficient quantum iO. Once such an efficient quantum iO primitive becomes available, the present construction can be instantiated by applying classical fully succinct iO to the GGM/PPRF component and quantum iO to the Hadamard-based signing component.

\section{Conclusions}
\label{sec:conclusions}

We have presented an explicit quantum-circuit implementation of a one-shot signature scheme, translating recent protocol-level constructions into concrete algorithms for $\textsc{KeyGen}$, $\textsc{Sign}$, and $\textsc{Verify}$. The construction prepares the quantum signing key as a uniform superposition over a hidden affine coset determined by a puncturable pseudorandom function, and realizes the required arithmetic using a $\mathrm{GGM}$-based lookup together with reversible linear-algebra subroutines. We also introduced a streamlined signing procedure for $l$-bit messages, replacing the repeated single-bit signing structure with a global measurement and a single translation step. Our resource analysis shows that the resulting implementation has logical qubit complexity scaling as $\Theta(\kappa \log r+n+l)$ and gate complexity scaling as $\Theta(n^3+nl)$, thereby making explicit the dominant costs and their dependence on the security and message-size parameters. From a security perspective, the construction should be viewed as a concrete algorithmic implementation of the unobfuscated $\mathrm{OSS}$ protocol, together with an identification of the components that must be hidden to obtain the intended one-shot guarantee. Classical obfuscation techniques can address the purely classical $\mathrm{PPRF}/\mathrm{GGM}$ subroutines, but the modified key generation step requires an obfuscator for the relevant class of quantum circuits. Thus, this work clarifies both the practical algorithmic structure of $\mathrm{OSS}$ and the remaining bottleneck for a full plain-model instantiation: the development of efficient and secure quantum obfuscation tools. We expect the circuit-level perspective developed here to be useful for future implementations of unclonable cryptographic primitives, including delegated signatures, quantum tokens, quantum money, and publicly verifiable randomness.

\section{Acknowledgements}
We acknowledge helpful discussions with Steven Duong.
\bibliography{main}

@misc{long-message-one-shot,
      title={A Simple and Efficient One-Shot Signature Scheme}, 
      author={Andrew Huang and Vinod Vaikuntanathan},
      year={2025},
      eprint={2510.10899},
      archivePrefix={arXiv},
      primaryClass={quant-ph},
      url={https://arxiv.org/abs/2510.10899}, 
}

@inproceedings{sz25,
author = {Shmueli, Omri and Zhandry, Mark},
title = {On One-Shot Signatures, Quantum vs. Classical Binding, and Obfuscating Permutations},
year = {2025},
isbn = {978-3-032-01877-9},
publisher = {Springer-Verlag},
address = {Berlin, Heidelberg},
url = {https://doi.org/10.1007/978-3-032-01878-6_12},
doi = {10.1007/978-3-032-01878-6_12},
booktitle = {Advances in Cryptology – CRYPTO 2025: 45th Annual International Cryptology Conference, Santa Barbara, CA, USA, August 17–21, 2025, Proceedings, Part II},
pages = {350–383},
numpages = {34},
location = {Santa Barbara, CA, USA}
}

@inproceedings{one-shot,
author = {Amos, Ryan and Georgiou, Marios and Kiayias, Aggelos and Zhandry, Mark},
title = {One-shot signatures and applications to hybrid quantum/classical authentication},
year = {2020},
isbn = {9781450369794},
publisher = {Association for Computing Machinery},
address = {New York, NY, USA},
url = {https://doi.org/10.1145/3357713.3384304},
doi = {10.1145/3357713.3384304},
booktitle = {Proceedings of the 52nd Annual ACM SIGACT Symposium on Theory of Computing},
pages = {255–268},
numpages = {14},
location = {Chicago, IL, USA},
series = {STOC 2020}
}

@article{GGM84,
author = {Goldreich, Oded and Goldwasser, Shafi and Micali, Silvio},
title = {How to construct random functions},
year = {1986},
issue_date = {Oct. 1986},
publisher = {Association for Computing Machinery},
address = {New York, NY, USA},
volume = {33},
number = {4},
issn = {0004-5411},
url = {https://doi.org/10.1145/6490.6503},
doi = {10.1145/6490.6503},
journal = {J. ACM},
month = aug,
pages = {792–807},
numpages = {16}
}

@misc{obfuscation,
  author       = {Mark Zhandry},
  title        = {Recent Developments in Program Obfuscation },
  year         = {2016},
  howpublished = {\url{https://mzhandry.github.io/courses/2016-Fall-COS597C/}},
  note         = {Accessed: 2025-06-24}
}

@article{bruhat_decomposion,
author = {Charles W. Curtis},
title = {{Representations of finite groups of Lie type}},
volume = {1},
journal = {Bulletin (New Series) of the American Mathematical Society},
number = {5},
publisher = {American Mathematical Society},
pages = {721 -- 757},
year = {1979},
}

@InProceedings{present,
author="Bogdanov, A.
and Knudsen, L. R.
and Leander, G.
and Paar, C.
and Poschmann, A.
and Robshaw, M. J. B.
and Seurin, Y.
and Vikkelsoe, C.",
editor="Paillier, Pascal
and Verbauwhede, Ingrid",
title="PRESENT: An Ultra-Lightweight Block Cipher",
booktitle="Cryptographic Hardware and Embedded Systems - CHES 2007",
year="2007",
publisher="Springer Berlin Heidelberg",
address="Berlin, Heidelberg",
pages="450--466",
isbn="978-3-540-74735-2"
}

@Article{present_implementation,
AUTHOR = {Jang, Kyungbae and Song, Gyeongju and Kim, Hyunjun and Kwon, Hyeokdong and Kim, Hyunji and Seo, Hwajeong},
TITLE = {Efficient Implementation of PRESENT and GIFT on Quantum Computers},
JOURNAL = {Applied Sciences},
VOLUME = {11},
YEAR = {2021},
NUMBER = {11},
ARTICLE-NUMBER = {4776},
URL = {https://www.mdpi.com/2076-3417/11/11/4776},
ISSN = {2076-3417},
DOI = {10.3390/app11114776}
}

@INPROCEEDINGS{SIMON,
  author={Beaulieu, Ray and Treatman-Clark, Stefan and Shors, Douglas and Weeks, Bryan and Smith, Jason and Wingers, Louis},
  booktitle={2015 52nd ACM/EDAC/IEEE Design Automation Conference (DAC)}, 
  title={The SIMON and SPECK lightweight block ciphers}, 
  year={2015},
  volume={},
  number={},
  pages={1-6},
  doi={10.1145/2744769.2747946}}

@article{grover_on_SIMON,
  title = {Grover on {$\,SIMON$}},
  author = {Ravi Anand and Arpita Maitra and Sourav Mukhopadhyay},
  journal = {Quantum Information Processing},
  year = {2020},
  volume = {19},
  number = {9},
  pages = {340},
  isbn = {1573-1332},
  doi = {10.1007/s11128-020-02844-w},
  url = {https://doi.org/10.1007/s11128-020-02844-w}
}

@article{iO,
author = {Barak, Boaz and Goldreich, Oded and Impagliazzo, Russell and Rudich, Steven and Sahai, Amit and Vadhan, Salil and Yang, Ke},
title = {On the (im)possibility of obfuscating programs},
year = {2012},
issue_date = {April 2012},
publisher = {Association for Computing Machinery},
address = {New York, NY, USA},
volume = {59},
number = {2},
issn = {0004-5411},
url = {https://doi.org/10.1145/2160158.2160159},
doi = {10.1145/2160158.2160159},
journal = {J. ACM},
month = may,
articleno = {6},
numpages = {48}
}

@misc{SZ25b,
      title={Unclonable Cryptography in Linear Quantum Memory}, 
      author={Omri Shmueli and Mark Zhandry},
      year={2025},
      eprint={2511.04633},
      archivePrefix={arXiv},
      primaryClass={quant-ph},
      url={https://arxiv.org/abs/2511.04633}, 
}

@article{BenDavid2023quantumtokens,
  doi = {10.22331/q-2023-01-19-901},
  url = {https://doi.org/10.22331/q-2023-01-19-901},
  title = {Quantum {T}okens for {D}igital {S}ignatures},
  author = {Ben-David, Shalev and Sattath, Or},
  journal = {{Quantum}},
  issn = {2521-327X},
  publisher = {{Verein zur F{\"{o}}rderung des Open Access Publizierens in den Quantenwissenschaften}},
  volume = {7},
  pages = {901},
  month = jan,
  year = {2023}
}

@article{Wiesner83,
author = {Wiesner, Stephen},
title = {Conjugate coding},
year = {1983},
issue_date = {Winter-Spring 1983},
publisher = {Association for Computing Machinery},
address = {New York, NY, USA},
volume = {15},
number = {1},
issn = {0163-5700},
url = {https://doi.org/10.1145/1008908.1008920},
doi = {10.1145/1008908.1008920},
journal = {SIGACT News},
month = jan,
pages = {78–88},
numpages = {11}
}

@inproceedings{AC12,
author = {Aaronson, Scott and Christiano, Paul},
title = {Quantum money from hidden subspaces},
year = {2012},
isbn = {9781450312455},
publisher = {Association for Computing Machinery},
address = {New York, NY, USA},
url = {https://doi.org/10.1145/2213977.2213983},
doi = {10.1145/2213977.2213983},
booktitle = {Proceedings of the Forty-Fourth Annual ACM Symposium on Theory of Computing},
pages = {41–60},
numpages = {20},
location = {New York, New York, USA},
series = {STOC '12}
}

@inproceedings{Shm22,
author = {Shmueli, Omri},
title = {Public-key Quantum money with a classical bank},
year = {2022},
isbn = {9781450392648},
publisher = {Association for Computing Machinery},
address = {New York, NY, USA},
url = {https://doi.org/10.1145/3519935.3519952},
doi = {10.1145/3519935.3519952},
booktitle = {Proceedings of the 54th Annual ACM SIGACT Symposium on Theory of Computing},
pages = {790–803},
numpages = {14},
location = {Rome, Italy},
series = {STOC 2022}
}

@article{quantum_lightning,
author = {Zhandry, Mark},
title = {Quantum Lightning Never Strikes the Same State Twice. Or: Quantum Money from Cryptographic Assumptions},
year = {2021},
issue_date = {Jan 2021},
publisher = {Springer-Verlag},
address = {Berlin, Heidelberg},
volume = {34},
number = {1},
issn = {0933-2790},
url = {https://doi.org/10.1007/s00145-020-09372-x},
doi = {10.1007/s00145-020-09372-x},
journal = {J. Cryptol.},
month = jan,
numpages = {56}
}

@misc{sattath2022prudent,
      title={Quantum Prudent Contracts with Applications to Bitcoin}, 
      author={Or Sattath},
      year={2022},
      eprint={2204.12806},
      archivePrefix={arXiv},
      primaryClass={quant-ph},
      url={https://arxiv.org/abs/2204.12806}, 
}

@article{coladangelo2020quantum,
   title={A Quantum Money Solution to the Blockchain Scalability Problem},
   volume={4},
   ISSN={2521-327X},
   url={http://dx.doi.org/10.22331/q-2020-07-16-297},
   DOI={10.22331/q-2020-07-16-297},
   journal={Quantum},
   publisher={Verein zur Forderung des Open Access Publizierens in den Quantenwissenschaften},
   author={Coladangelo, Andrea and Sattath, Or},
   year={2020},
   month={July}, pages={297} }

@misc{quasilineriO,
      author = {Yaohua Ma and Chenxin Dai and Elaine Shi},
      title = {Quasi-Linear Indistinguishability Obfuscation via Mathematical Proofs of Equivalence and Applications},
      howpublished = {Cryptology {ePrint} Archive, Paper 2025/307},
      year = {2025},
      url = {https://eprint.iacr.org/2025/307}
}

@misc{PrO,
      author = {Pedro Branco and Nico Döttling and Abhishek Jain and Giulio Malavolta and Surya Mathialagan and Spencer Peters and Vinod Vaikuntanathan},
      title = {Pseudorandom Obfuscation and Applications},
      howpublished = {Cryptology {ePrint} Archive, Paper 2024/1742},
      year = {2024},
      url = {https://eprint.iacr.org/2024/1742}
}

@misc{diamondiO,
      author = {Sora Suegami and Enrico Bottazzi and Gayeong Park},
      title = {Diamond {iO}: A Straightforward Construction of Indistinguishability Obfuscation from Lattices},
      howpublished = {Cryptology {ePrint} Archive, Paper 2025/236},
      year = {2025},
      url = {https://eprint.iacr.org/2025/236}
}

@INPROCEEDINGS {pviO,
author = { Jain, Abhishek and Jin, Zhengzhong and Mathialagan, Surya and Paneth, Omer },
booktitle = { 2025 IEEE 66th Annual Symposium on Foundations of Computer Science (FOCS) },
title = {{ On Succinct Obfuscation via Propositional Proofs }},
year = {2025},
volume = {},
ISSN = {},
pages = {1703-1740},
doi = {10.1109/FOCS63196.2025.00091},
url = {https://doi.ieeecomputersociety.org/10.1109/FOCS63196.2025.00091},
publisher = {IEEE Computer Society},
address = {Los Alamitos, CA, USA},
month =Dec}

@misc{obfusqate,
      title={ObfusQate: Unveiling the First Quantum Program Obfuscation Framework}, 
      author={Nikhil Bartake and See Toh Zi Jie and Carmen Wong Jiawen and Michael Kasper and Vivek Balachandran},
      year={2025},
      eprint={2503.23785},
      archivePrefix={arXiv},
      primaryClass={cs.CR},
      url={https://arxiv.org/abs/2503.23785}, 
}

@inproceedings{Opaque,
author = {Patel, Tirthak and Ranjan, Aditya and Silver, Daniel and Gandhi, Harshitta and Cutler, William and Tiwari, Devesh},
title = {OpaQue: Program Output Obfuscation for Quantum Software Circuits in Quantum Clouds},
year = {2025},
isbn = {9798400715372},
publisher = {Association for Computing Machinery},
address = {New York, NY, USA},
url = {https://doi.org/10.1145/3721145.3725771},
doi = {10.1145/3721145.3725771},
booktitle = {Proceedings of the 39th ACM International Conference on Supercomputing},
pages = {1079–1091},
numpages = {13},
location = {
},
series = {ICS '25}
}

@INPROCEEDINGS{attackOpaque,
  author={Lushi, Donald and Saeed, Samah Mohamed},
  booktitle={2025 IEEE International Symposium on Hardware Oriented Security and Trust (HOST)}, 
  title={Undermining Quantum Circuit Obfuscation: Insights from Structural Analysis}, 
  year={2025},
  volume={},
  number={},
  pages={88-98},
  doi={10.1109/HOST64725.2025.11050070}}

@inproceedings{GR07,
  title={On best-possible obfuscation},
  author={Goldwasser, Shafi and Rothblum, Guy N},
  booktitle={Theory of Cryptography Conference},
  pages={194--213},
  year={2007},
  organization={Springer}
}

@article{GGHRS+16,
  title={Candidate indistinguishability obfuscation and functional encryption for all circuits},
  author={Garg, Sanjam and Gentry, Craig and Halevi, Shai and Raykova, Mariana and Sahai, Amit and Waters, Brent},
  journal={SIAM Journal on Computing},
  volume={45},
  number={3},
  pages={882--929},
  year={2016},
  publisher={SIAM}
}

@inproceedings{ADYS14,
  title={Optimizing obfuscation: avoiding Barrington's theorem},
  author={Ananth, Prabhanjan and Gupta, Divya and Ishai, Yuval and Sahai, Amit},
  booktitle={Proceedings of the 2014 ACM SIGSAC Conference on Computer and Communications Security},
  pages={646--658},
  year={2014}
}

@article{BV18,
  title={Indistinguishability obfuscation from functional encryption},
  author={Bitansky, Nir and Vaikuntanathan, Vinod},
  journal={Journal of the ACM (JACM)},
  volume={65},
  number={6},
  pages={1--37},
  year={2018},
  publisher={ACM New York, NY, USA}
}

@inproceedings{AJS17,
  title={Indistinguishability obfuscation for turing machines: constant overhead and amortization},
  author={Ananth, Prabhanjan and Jain, Abhishek and Sahai, Amit},
  booktitle={Annual International Cryptology Conference},
  pages={252--279},
  year={2017},
  organization={Springer}
}

@article{JLS26,
  title={Indistinguishability obfuscation from well-founded assumptions},
  author={Jain, Aayush and Lin, Huijia and Sahai, Amit},
  journal={Journal of the ACM},
  volume={73},
  number={1},
  pages={1--30},
  year={2026},
  publisher={ACM New York, NY}
}

@inproceedings{JJ22,
  title={Indistinguishability obfuscation via mathematical proofs of equivalence},
  author={Jain, Abhishek and Jin, Zhengzhong},
  booktitle={2022 IEEE 63rd Annual Symposium on Foundations of Computer Science (FOCS)},
  pages={1023--1034},
  year={2022},
  organization={IEEE}
}

@inproceedings{MDS25,
  title={Quasi-linear indistinguishability obfuscation via mathematical proofs of equivalence and applications},
  author={Ma, Yaohua and Dai, Chenxin and Shi, Elaine},
  booktitle={Annual International Conference on the Theory and Applications of Cryptographic Techniques},
  pages={157--186},
  year={2025},
  organization={Springer}
}

@inproceedings{RVV25,
  title={Indistinguishability obfuscation from bilinear maps and LPN variants},
  author={Ragavan, Seyoon and Vafa, Neekon and Vaikuntanathan, Vinod},
  booktitle={Theory of Cryptography Conference},
  pages={3--36},
  year={2025},
  organization={Springer}
}

@article{AF16,
  title={On quantum obfuscation},
  author={Alagic, Gorjan and Fefferman, Bill},
  journal={arXiv preprint arXiv:1602.01771},
  year={2016}
}

@inproceedings{ABDS21,
  title={Impossibility of quantum virtual black-box obfuscation of classical circuits},
  author={Alagic, Gorjan and Brakerski, Zvika and Dulek, Yfke and Schaffner, Christian},
  booktitle={Annual International Cryptology Conference},
  pages={497--525},
  year={2021},
  organization={Springer}
}

@inproceedings{JJMP25,
  title={On succinct obfuscation via propositional proofs},
  author={Jain, Abhishek and Jin, Zhengzhong and Mathialagan, Surya and Paneth, Omer},
  booktitle={2025 IEEE 66th Annual Symposium on Foundations of Computer Science (FOCS)},
  pages={1703--1740},
  year={2025},
  organization={IEEE}
}

@inproceedings{BK21,
  title={Constructions for quantum indistinguishability obfuscation},
  author={Broadbent, Anne and Kazmi, Raza Ali},
  booktitle={International Conference on Cryptology and Information Security in Latin America},
  pages={24--43},
  year={2021},
  organization={Springer}
}

@inproceedings{BKNY23,
  title={Obfuscation of pseudo-deterministic quantum circuits},
  author={Bartusek, James and Kitagawa, Fuyuki and Nishimaki, Ryo and Yamakawa, Takashi},
  booktitle={Proceedings of the 55th Annual ACM Symposium on Theory of Computing},
  pages={1567--1578},
  year={2023}
}

@inproceedings{CG24,
  title={How to use quantum indistinguishability obfuscation},
  author={Coladangelo, Andrea and Gunn, Sam},
  booktitle={Proceedings of the 56th Annual ACM Symposium on Theory of Computing},
  pages={1003--1008},
  year={2024}
}

@inproceedings{BBV24,
  title={Quantum state obfuscation from classical oracles},
  author={Bartusek, James and Brakerski, Zvika and Vaikuntanathan, Vinod},
  booktitle={Proceedings of the 56th Annual ACM Symposium on Theory of Computing},
  pages={1009--1017},
  year={2024}
}

@inproceedings{HT25,
  title={Obfuscation of unitary quantum programs},
  author={Huang, Mi-Ying Miryam and Tang, Er-Cheng},
  booktitle={2025 IEEE 66th Annual Symposium on Foundations of Computer Science (FOCS)},
  pages={1665--1671},
  year={2025},
  organization={IEEE}
}

@article{HT26,
  title={Obfuscation of Arbitrary Quantum Circuits},
  author={Huang, Miryam Mi-Ying and Tang, Er-Cheng},
  journal={arXiv preprint arXiv:2601.08969},
  year={2026}
}

@inproceedings{shor1994,
  title={Algorithms for quantum computation: discrete logarithms and factoring},
  author={Shor, Peter W},
  booktitle={Proceedings 35th annual symposium on foundations of computer science},
  pages={124--134},
  year={1994},
  organization={Ieee}
}

@article{shor1999,
  title={Polynomial-time algorithms for prime factorization and discrete logarithms on a quantum computer},
  author={Shor, Peter W},
  journal={SIAM review},
  volume={41},
  number={2},
  pages={303--332},
  year={1999},
  publisher={SIAM}
}

@inproceedings{grover1996,
  title={A fast quantum mechanical algorithm for database search},
  author={Grover, Lov K},
  booktitle={Proceedings of the twenty-eighth annual ACM symposium on Theory of computing},
  pages={212--219},
  year={1996}
}

@article{NIST_PQC,
  title={Report on post-quantum cryptography},
  author={Chen, Lily and Chen, Stephen and Jordan, Stephen and Liu, Yi-Kai and Moody, Dustin and Peralta, Rene and Perlner, Ray and Smith-Tone, Daniel},
  journal={National Institute of Standards and Technology Internal Report},
  volume={8105},
  pages={1--15},
  year={2016}
}

@article{wootters1982single,
  title={A single quantum cannot be cloned},
  author={Wootters, William K and Zurek, Wojciech H},
  journal={Nature},
  volume={299},
  number={5886},
  pages={802--803},
  year={1982},
  publisher={Nature Publishing Group}
}

@article{bennett89,
author = {Bennett, Charles H.},
title = {Time/space trade-offs for reversible computation},
year = {1989},
issue_date = {Aug. 1989},
publisher = {Society for Industrial and Applied Mathematics},
address = {USA},
volume = {18},
number = {4},
issn = {0097-5397},
url = {https://doi.org/10.1137/0218053},
doi = {10.1137/0218053},
journal = {SIAM J. Comput.},
month = aug,
pages = {766–776},
numpages = {11}
}

@article{cakan2025public,
  title={Public-key quantum fire and key-fire from classical oracles},
  author={Cakan, Alper and Goyal, Vipul and Shmueli, Omri},
  journal={arXiv preprint arXiv:2504.16407},
  year={2025}
}

@inproceedings{bostanci2025general,
  title={A general quantum duality for representations of groups with applications to quantum money, lightning, and fire},
  author={Bostanci, John and Nehoran, Barak and Zhandry, Mark},
  booktitle={Proceedings of the 57th Annual ACM Symposium on Theory of Computing},
  pages={201--212},
  year={2025}
}

@article{casper2026publicly,
  title={Publicly Certifiable Min-Entropy Without Quantum Communication},
  author={Casper, Ofer and Nehoran, Barak and Sattath, Or},
  journal={Cryptology ePrint Archive},
  year={2026}
}

@inproceedings{Cook75,
  author    = {Stephen A. Cook},
  title     = {Feasibly Constructive Proofs and the Propositional Calculus},
  booktitle = {Proceedings of the Seventh Annual ACM Symposium on Theory of Computing},
  pages     = {83--97},
  publisher = {ACM},
  year      = {1975}
}

@article{CookReckhow79,
  author  = {Stephen A. Cook and Robert A. Reckhow},
  title   = {The Relative Efficiency of Propositional Proof Systems},
  journal = {Journal of Symbolic Logic},
  volume  = {44},
  number  = {1},
  pages   = {36--50},
  year    = {1979}
}

@misc{CLLZ22,
      title={Hidden Cosets and Applications to Unclonable Cryptography}, 
      author={Andrea Coladangelo and Jiahui Liu and Qipeng Liu and Mark Zhandry},
      year={2022},
      eprint={2107.05692},
      archivePrefix={arXiv},
      primaryClass={cs.CR},
      url={https://arxiv.org/abs/2107.05692}, 
}

@book{katz2007,
  title={Introduction to modern cryptography: principles and protocols},
  author={Katz, Jonathan and Lindell, Yehuda},
  year={2007},
  publisher={Chapman and hall/CRC}
}

@misc{puncturedprograms13,
      author = {Amit Sahai and Brent Waters},
      title = {How to Use Indistinguishability Obfuscation: Deniable Encryption, and More},
      howpublished = {Cryptology {ePrint} Archive, Paper 2013/454},
      year = {2013},
      url = {https://eprint.iacr.org/2013/454}
}

@article{Khattar2025riseofconditionally,
  doi = {10.22331/q-2025-05-21-1752},
  url = {https://doi.org/10.22331/q-2025-05-21-1752},
  title = {Rise of conditionally clean ancillae for efficient quantum circuit constructions},
  author = {Khattar, Tanuj and Gidney, Craig},
  journal = {{Quantum}},
  issn = {2521-327X},
  publisher = {{Verein zur F{\"{o}}rderung des Open Access Publizierens in den Quantenwissenschaften}},
  volume = {9},
  pages = {1752},
  month = may,
  year = {2025}}

@article{Mallows2,
title = {The two-sided infinite extension of the Mallows model for random permutations},
journal = {Advances in Applied Mathematics},
volume = {48},
number = {5},
pages = {615-639},
year = {2012},
issn = {0196-8858},
doi = {https://doi.org/10.1016/j.aam.2012.01.001},
url = {https://www.sciencedirect.com/science/article/pii/S0196885812000115},
author = {Alexander Gnedin and Grigori Olshanski}
}

@article{Mallows1,
    author = {MALLOWS, C. L.},
    title = {NON-NULL RANKING MODELS. I},
    journal = {Biometrika},
    volume = {44},
    number = {1-2},
    pages = {114-130},
    year = {1957},
    month = {06},
    issn = {0006-3444},
    doi = {10.1093/biomet/44.1-2.114},
    url = {https://doi.org/10.1093/biomet/44.1-2.114},
    eprint = {https://academic.oup.com/biomet/article-pdf/44/1-2/114/752590/44-1-2-114.pdf},
}
\bibliographystyle{alpha}

\end{document}